\documentclass[twocolumn,notitlepage,aps,prb,10pt,superscriptaddress,floatfix,letterpaper,reprint
]{revtex4-2}

\usepackage{hyperref}
\usepackage[usenames,dvipsnames]{color}
\usepackage{amsmath}
\usepackage[english]{babel}
\usepackage{amssymb}
\usepackage{graphicx}
\usepackage{epstopdf}
\usepackage[normalem]{ulem}
\usepackage{subfigure}
\usepackage{todonotes}
\usepackage{braket}
\usepackage{bbold}
\usepackage{placeins}
\usepackage[para]{threeparttable}
\usepackage{lipsum}
\usepackage{multirow}
\usepackage{bm}
\usepackage{xcolor}
\usepackage{amsfonts}
\definecolor{linkcolor}{HTML}{399B03}
\definecolor{urlcolor}{HTML}{399B03}
\hypersetup{pdfstartview=FitH, linkcolor=linkcolor,urlcolor=urlcolor,colorlinks=true}
\hypersetup{
    unicode=false,          
    pdftoolbar=true,        
    pdfmenubar=true,        
    pdffitwindow=false,     
    pdfstartview={FitH},    
    pdfauthor={},         
    colorlinks=true,        
    linkcolor=blue,         
    citecolor=red,          
}

\newcommand{\tk}{\textbf{k}}
\newcommand{\tq}{\textbf{q}}
\newcommand{\tK}{\textbf{K}}
\newcommand{\tQ}{\textbf{Q}}

\newcommand{\tr}{\textbf{r}}

\makeatletter
\newcommand*{\balancecolsandclearpage}{%
    \close@column@grid
    \cleardoublepage
    \twocolumngrid
}
\makeatother
\begin{document}

 \title{Single- and two-particle finite size effects in interacting lattice systems}

\author{Sergei Iskakov}
\affiliation{Department of Physics, University of Michigan, Ann Arbor, MI 48109, USA}
\author{Hanna Terletska}
\affiliation{Department of Physics and Astronomy, Middle Tennessee State University, Murfreesboro, TN 37123, USA}
\author{Emanuel Gull}
\affiliation{Department of Physics, University of Michigan, Ann Arbor, MI 48109, USA}
\date{\today}

\begin{abstract}
Simulations of extended quantum systems are typically performed by extrapolating results of a sequence of finite-system-size simulations to the thermodynamic limit. In the quantum Monte Carlo community,  twist-averaging was pioneered as an efficient strategy to eliminate one-body finite size effects. In the dynamical mean field community, cluster generalizations of the dynamical mean field theory were formulated to study  systems with non-local correlations. In this work, we put the twist-averaging and the dynamical cluster approximation variant of the dynamical mean field theory onto equal footing, discuss commonalities and differences, and compare results from both techniques to the standard periodic boundary technique. At the example of Hubbard-type models with local, short-range and Yukawa-like longer range interactions we show that all methods converge to the same limit, but that the convergence speed  differs in practice. We show that embedding theories are an effective tool for managing both one-body and two-body finite size effects, in particular if interactions are averaged over twist angles.
\end{abstract}

\maketitle
\section{Introduction}\label{sec:introduction}
Understanding the physics of periodic solids and periodic lattice model systems requires describing their properties in the `thermodynamic' or macroscopic limit of infinite system size.
Most numerical methods approach this limit with results for a sequence of successively larger  finite-size systems, which are then extrapolated to the thermodynamic limit using a known scaling behavior. Since the numerical effort grows either polynomially or exponentially with system size, constructing reliable finite size extrapolations from small systems is important.

Approximation errors may result from the truncation of the Hamiltonian to the finite system and from the introduction of an artificial boundary or surface. Periodic boundary conditions (PBC), which are the standard choice for the simulation of periodic systems, eliminate boundary effects. 
Finite size effects, which result from the truncation of the Hamiltonian and its solution to the finite system (or, equivalently, to a discrete set of momentum points in the Brillouin zone), remain. Two strategies to overcome such effects have been proposed.

First, one may introduce a twist to the boundary conditions, resulting in a shift of the corresponding momentum points~\cite{Valenti1991,Gros1992,Gros1996,Ceperley2001}. In the calculation of local observables, simulations for multiple `twisted' systems are then averaged over twist angles. This `twist-averaged boundary condition' (TABC) technique is predominantly used in Monte Carlo calculations, since the average can be taken during the simulation at no additional cost.

Second, one may construct a quantum embedding method. This is the rationale behind cluster dynamical mean field methods~\cite{Hettler1998,Lichtenstein2000,Kotliar2001,Maier2005} such as the dynamical cluster approximation (DCA)~\cite{Hettler1998,Maier2005}, which constructs a periodic embedding. DCA can be viewed as a generalization dynamical mean field theory (DMFT) to `patches' in momentum space, where propagators, interactions, and correlations are chosen to be identical in a patch in momentum space~\cite{Fuhrmann2007} but correlations retain their full frequency dependence. The numerical effort of this technique is typically $\sim10\times$ the effort of the PBC solution, since the DCA equations have to be solved self-consistently. In self-consistent diagrammatic calculations, the DCA and the diagrammatic self-consistency can be performed concurrently at no additional cost.

Twist-averaging and DCA embedding are remarkably similar. Both perform an average over shifted momentum points, and both have the same asymptotic convergence scaling, since for local observables and local interactions, all of these strategies converge  to the exact result with a quadratic convergence in the linear system size but with different prefactors~\cite{Hettler1998,Fuchs2011} (faster convergence may be observed under special circumstances~\cite{Biroli2002}). They differ in the place where the average is taken, which has consequences for the numerical solution methods and for the prefactor of the convergence speed.

In this paper, we examine the convergence of these three strategies (PBC, DCA and TABC). We show that, as expected, they converge to the same solution in practice, and we examine convergence speed for different types of finite size effects at the example of three different lattice models. In particular, we examine the case of longer range interactions and highlight the importance of averaging interactions in addition to the one-body Hamiltonian in such systems.

The remainder of this paper is as follows. In Sec.~\ref{sec:methods} we introduce the models studied, the finite size correction methods, and the solution methodology. In Sec.~\ref{sec:results} we present results, and Sec.~\ref{sec:conclusions} contains conclusions.

\section{Model and Methods}\label{sec:methods}
\subsection{Hubbard models with local and non-local interactions}\label{subsec:models}
We illustrate finite size effects at the example of the fermionic single-band Hubbard-like Hamiltonian~\cite{Qin2022,Arovas2022}
with non-local density-density interactions,
    \begin{align}
        H = -t \sum_{\left<i,j\right>,\sigma} c^{\dagger}_{i,\sigma} c_{j,\sigma}
        +&\frac{1}{2}\sum_{i,j,\sigma\sigma'} \tilde{U}_{ij} n_{i,\sigma}n_{j,\sigma'}.
        \label{eqn:Ham}
    \end{align}
 $\sigma$ labels the spin index,
 $i$ and $j$ are site indices, and $\left<i, j\right>$ denotes a sum over nearest neighbors. $c^{\dagger}_{i,\sigma}$ ($c_{j,
 \sigma}$) are creation (annihilation) operators and $n_{i,\sigma} = c^{\dagger}_{i,\sigma}c_{i,\sigma}$ describe the
 particle number operators. $t$ denotes the nearest-neighbor hopping and $\tilde{U}_{ij}$ the density-density interaction
 strength.  We write $H=H_0+V$, where $H_0$ denotes the quadratic hopping terms and $V$ denotes the interactions.
 We restrict ourselves to the model on a three-dimensional simple cubic lattice.

    We separate the interaction  into two parts,
    \begin{align}
        \tilde{U}_{ij} = U\delta_{ij} + V_{ij},
    \end{align}
    where $U$ is the onsite Hubbard interaction and $V_{ij}$ the non-local part, and discuss three different choices of $V_{ij}$.
    
For  $V_{ij} \equiv 0$ we obtain the standard three-dimensional Hubbard model. Due to its relevance to cold atomic gases \cite{Hart2015} and a second-order phase transition at high temperature, this model has become a fruitful testbed for many-body methods and numerically exact results are available from several reliable techniques \cite{Staudt2000,Kent2005,Fuchs2011c,Fuchs2011,DeLeo2011,Paiva2011,Kozik2013,Duarte2015,PhysRevB.105.045109}, especially for interactions near the maximal critical temperature.
    
For
    \begin{align}
        V_{ij} = V_{12}\delta_{\langle ij\rangle},
    \end{align}
    with $V_{12}$ the nearest neighbor interaction strength and $\delta_{\langle ij\rangle}=1$ if $i$ and $j$ are neighbors, and zero otherwise, we obtain the extended Hubbard model. The nearest-neighbor interactions in this model induce a
 charge-order phase transition which has been extensively studied with DCA~\cite{Terletska2017,Terletska2018,Terletska2019,Terletska2021},
 extended DMFT~\cite{PhysRevLett.84.3678,Camjayi2008,PhysRevB.82.155102,PhysRevB.95.125112},
 DMFT+GW~\cite{PhysRevB.66.085120,PhysRevLett.109.226401,PhysRevB.87.125149,PhysRevB.90.195114}, and dual boson
 techniques~\cite{Rubtsov2012,PhysRevB.90.235135,PhysRevB.93.045107,PhysRevB.100.165128,PhysRevB.102.195109}.
    
For 
    \begin{align}
        V_{ij} = V_{0}\frac{e^{-\alpha |\tr_{ij}|}}{|\tr_{ij}|}\label{eq:Yukawa}
    \end{align}
we obtain a Yukawa Hubbard model. $V_{0} = \frac{V_{12}}{e^{-\alpha}}$ denotes the interaction strength, $\alpha$ the Yukawa decay constant,
    and $\tr_{ij} = \tr_i - \tr_j$ the real space distance between site $i$ and $j$. This model is chosen to mimic aspects of the long-range interactions occurring in electronic structure problems without introducing the complication of a divergent long-range potential.
    
In momentum space, we write
    \begin{align}
        \tilde{U}(\tq) &= U + V(\tq),\\
        V(\tq) &= \sum_{i,j} V_{ij} e^{-i\tq \tr_{ij}},
        \label{eqn:InteractionFourier}
    \end{align}
    with $\tq$ the bosonic momentum transfer vector. Note that general four-fermion interactions, which we do not study in this
 work, depend on three momenta and four unit cell indices.
    
We discuss results in the finite-temperature Matsubara formalism, where single-particle quantities can be expressed in terms of Matsubara Green's functions, self-energies, and interaction vertices. Matsubara Green's functions are given as \cite{Mahan2000}
\begin{align}
        G(\omega_n, \tk) = \left[ (i\omega_n +\mu) \mathbb{1} - F_{\tk} - {\Sigma}(\omega_n, \tk) \right]^{-1}.
   \label{eqn:G}
\end{align}
Here $\omega_n=(2n+1)\pi T$ denotes Matsubara frequencies at temperature $T$, $\mu$ the chemical potential, $F_\tk = H^{0}_\tk + \Sigma^{\infty}(\tk)$ the Fock matrix and $\Sigma$ the self-energy, which is uniquely defined if $F_\tk$ is chosen such that $\Sigma(i\omega_n\rightarrow i\infty)\rightarrow 0.$ The exact self-energy $\Sigma$ is a functional of the momentum-dependent Green's function and the momentum-dependent interaction and, in diagrammatic theories, is obtained by adding all so-called skeleton diagrams to infinite order \cite{Luttinger1960,Mahan2000}. Approximations, such as the $GW$ approximation \cite{Hedin1965}, can be constructed by considering a subset of all diagrams.

\subsection{Finite size effects}\label{subsec:fseffects}
The approximation of an infinite periodic system by an auxiliary finite size system introduces finite size approximations to the Hamiltonian $H=H_0+V$. 
The approximation of the non-interacting Hamiltonian $H_0$ on a discrete set of momentum points
 results in so-called `independent-particle', `single-particle', or `one-body' finite size effects~\cite{Kent1999,Drummond2008,Azadi2015,Dagrada2016}.

The  situation for the interaction $V$ is more complex. The Hubbard interaction is already constant in momentum space (local in real space) and therefore remains exact when approximated on a finite system.
The extended Hubbard interaction is short-range and captured exactly on all but the smallest systems.
In contrast, the Yukawa interaction of Eq.~\ref{eq:Yukawa} will become approximate when truncated to the size of a finite system; the
 approximation leads to a non-divergent but non-zero correction. In more general systems with long-ranged interactions, such as those generally considered in the electronic structure problem, the truncation of the interaction to a finite system may lead to a divergent contribution which needs to be carefully compensated~\cite{Fraser1996,Foulkes1997,Kent1999}, and which may dominate the finite size convergence~\cite{Wilhelm2017}.

In the following, we denote continuous momenta in the infinite system by $\tk,\tk',$ and $\tq$ and discrete momenta in the finite auxiliary system by $\tK, \tK'$, and $\tQ$, 

A finite size approximation $G(\omega_n, \tk) \rightarrow G(\omega_n, \tK)$ then introduces `one-body' finite size errors via the discretization of the $H^{(0)}_\tK$ contribution in the Fock matrix on a finite number of $\tK$-points, along with approximations to the interaction term. These approximations result in finite-size errors in the self-energy contributions $\Sigma^{\infty}(\tK)$ and $\Sigma(\omega_n,\tK)$, which  are generated from the approximations of $H_0$ and $V$ in a highly non-linear manner.

\begin{figure}[tbh]
    \includegraphics[width=0.97\columnwidth]{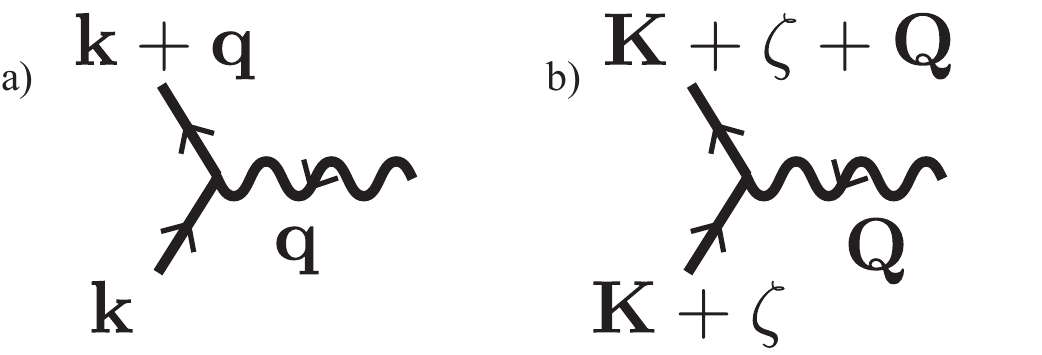}
    \caption{Left panel: Scattering vertex, scattering an electron of momentum $\tk$ to momentum $\tk+\tq$ while absorbing
    momentum $\tq$. Right panel: `twisted' version of the same vertex with twist $\zeta$, leaving the bosonic momentum invariant.}
       \label{fig:momcons_tabc}
\end{figure}

\subsection{Finite size corrections}\label{subsec:fscor}
\subsubsection{Standard methodology}
We start our discussion from local observables, such as the energy, the double-occupancy or the density. These quantities can be expressed as integrals over the Brillouin zone.

Most many-body methods are restricted to uniform momentum grids with regularly spaced momenta $\tK, \tK'$ and $\tQ$, since all incoming and outgoing momenta of a scattering vertex (see Fig.~\ref{fig:momcons_tabc}) have to lie on the momentum grid. Integrals of smooth functions over such regularly spaced, periodic grids with uniform weights converge quadratically in the number of points in each dimension, leading to a quadratic convergence of Brillouin zone integrals and thus to a quadratic convergence of local quantities. The exception to this rule are long-range interactions with a singularity at the $\Gamma$ point, see {\it e.g.} Refs.~\cite{Fraser1996,Foulkes1997,Kent1999,Wilhelm2017}.

In the standard methodology, simulations are performed on regularly spaced Monkhorst-Pack grids \cite{MonkhorstPack1976} which, for $\tQ$, include the $\Gamma$ point with zero momentum transfer. The  sum $\tK+\tQ$ of a fermionic momentum $\tK$ and a bosonic momentum $\tQ$ then lies on the fermionic grid, such that momentum conservation can be respected.

Local quantities are evaluated as a sum over the discrete $\tK$-points. Momentum-dependent quantities are only known at the discrete $\tK$ points. If a more precise momentum resolution is desired, additional momentum points have to be obtained via interpolation. In this work, momentum-dependent
quantities, such as the momentum-dependent Matsubara Green's function, are evaluated by performing a Wannier
interpolation~\cite{Wannier_RMP_2021} of the self-energy, from which the Green's function is recovered with an exactly evaluated non-interacting part of the Hamiltonian. We note that interplations of the so-called cumulant~\cite{Stanescu2006} and of the Green's function are also possible, along with spline interpolations of the self-energy \cite{Macridin06}. The interpolation scheme can have a large effect on the results if systems are small and self-energies are strongly momentum dependent \cite{Sakai2009,Sakai2009B,Sakai12}. However, we find that, in the systems analyzed here, results become independent of the interpolation scheme as the system size is sufficiently increased.

One may also break some of the lattice symmetries, rather than performing simulations on lattices that respect the full space-group symmetry of the infinite system,  and symmetrize results. Breaking symmetries gives access to additional finite systems, which can then be used to perform finite size extrapolations from a dense set of points. The technique was pioneered in the context of exact diagonalization \cite{Betts1997,Betts1999} and is standard in cluster dynamical mean field theory \cite{Maier2005} (see Ref.~\cite{PhysRevB.105.045109} for detailed results for the systems analyzed here).

\subsubsection{Twisted boundary conditions and twist averaging}\label{subsubsec:twist-averaged-boundary-conditions}
While the momenta $\tK, \tK'$ and $\tQ$ have to be on momentum grids with regularly spaced momenta, it is possible to shift the `fermionic' momentum grids for $\tK$ and $\tK'$ by an arbitrary momentum twist $\zeta$ while leaving the `bosonic' one for $\tQ$ invariant. The difference of two fermionic momenta $\tK+\zeta,\tK'+\zeta$ then still lies on the bosonic grid $\tQ$, and the sum of fermionic and bosonic momenta lies on the fermionic grid, such that momentum conservation can be respected. The resulting shifted scattering vertex is illustrated in the right panel of Fig.~\ref{fig:momcons_tabc}.

Momentum points can be shifted by introducing a phase factor or `twist' into the wave function or the hopping, and the technique of performing simulations for many twists and averaging the results leads to the twist-averaged boundary conditions~\cite{Gros1996,Ceperley2001}. Twist averaging is commonly used in
QMC simulations~\cite{Kent1999,Drummond2008,Drummond2013}, where twists are either chosen  randomly~\cite{Ceperley2001} or with uniformly spaced twist angles~\cite{Gros1996}.

Each twist $\zeta$ samples a different part of the one-body Hamiltonian $H^{0}_{\tK+\zeta}$, while the two-body interaction $V(\tq)$ remains approximated on the original $\tQ$-grid, $V(\tQ)$. Observables are evaluated for
each twist angle, and the corresponding local quantities are obtained by averaging over all twists.
In Monte Carlo methods, the average over twists is often obtained by sampling over $\zeta$, thereby incurring no additional cost.
In deterministic methods, or if momentum-dependent quantities are desired, the numerical effort increases proportionally to the number of twists.

In terms of a diagrammatic skeleton series, the relation between twisted propagators $G^t$, self-energies $\Sigma^t$, and interactions is
\begin{align}
G^t(\omega_n,\tK+\zeta)=[(i\omega_n + \mu)\mathbb{1} -F_{\tK+\zeta}-\Sigma^t(\omega_n, \tK+\zeta)]^{-1} \label{eq:twistedgf}
\end{align}
where $\Sigma^t(\omega_n,\tK+\zeta)=\Sigma^t[G^t(\omega_n, \tK+\zeta), U+V(\tQ)]$ is a functional of the twisted interacting Green's function for a given twist $\zeta$ and the interactions evaluated at momentum $\tQ$.

 \begin{figure*}[tb]
     \includegraphics[width=0.24\textwidth]{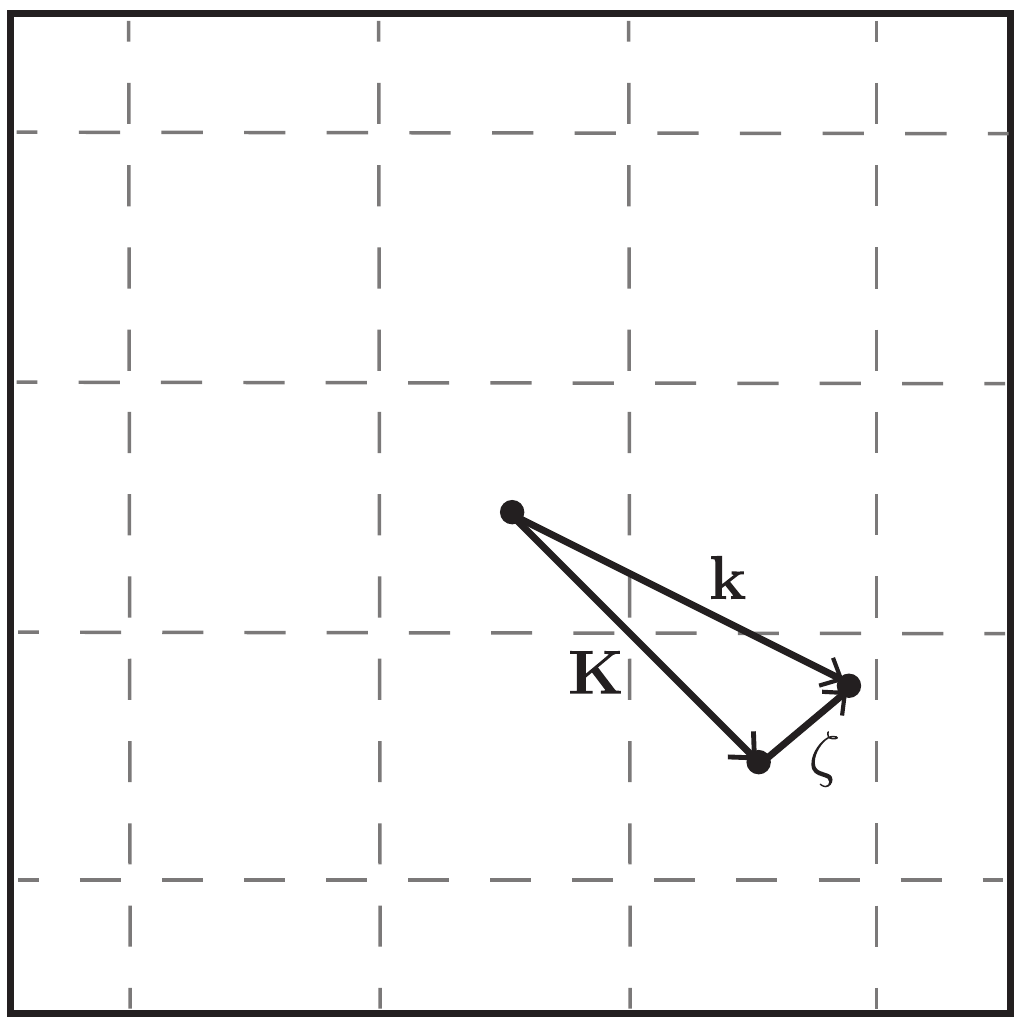}
     \includegraphics[width=0.74\textwidth]{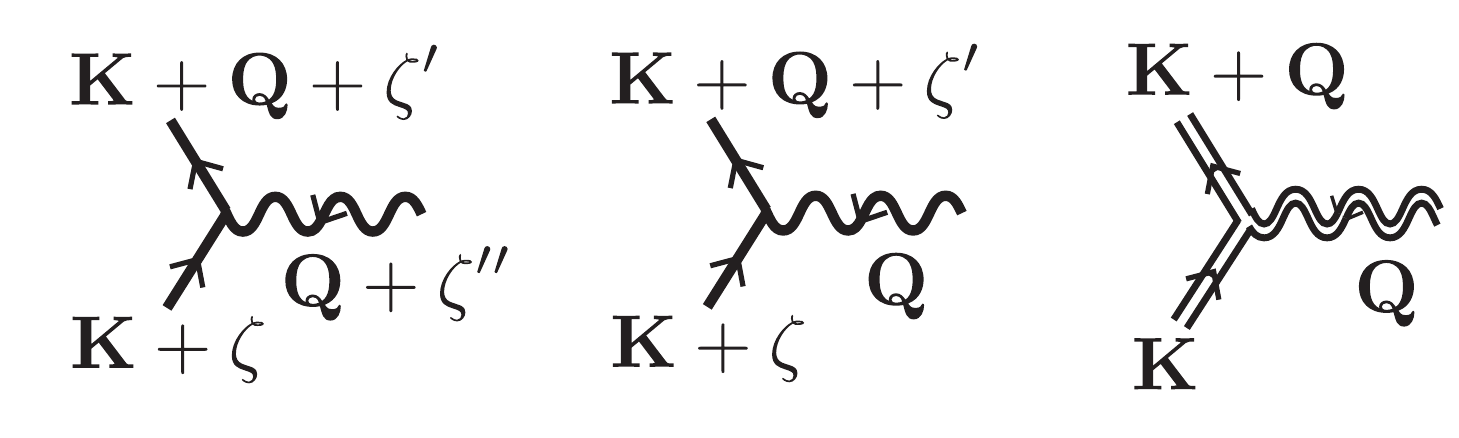}
     \caption{From left to right. First panel: DCA approximation scheme in the Brillouin zone of the 2D square lattice. An arbitrary lattice momentum $\tk$ is
     represented as the sum of a momentum $\zeta$ within a patch and a corresponding cluster momentum $\tK$ to the center of the patch.
Second panel: Scattering vertex in DCA with interactions $V^{c,a}$, scattering an electron of momentum $\tK+\zeta$ to momentum $\tK+\tQ+\zeta'$ while absorbing a momentum $\tQ+\zeta''$, violating momentum conservation on a scale comparable to the patch size. Third panel: same, for interactions $V^{c,i}$ with momentum $\tQ$. Fourth panel: cluster scattering diagram, respecting cluster momentum conservation. Straight lines denote lattice Green's functions with momenta indicated, wiggly lines momentum transfer, and vertices the interaction vertex. Double lines denote cluster quantities.}
     \label{fig:DCA}
 \end{figure*}

\subsubsection{Dynamical Cluster Approximation}\label{subsubsec:dca}
The Dynamical Cluster Approximation~\cite{Hettler1998,Maier2005} was introduced as a cluster generalization~\cite{Hettler1998,Kotliar2001,Lichtenstein2000} of the single-site DMFT~\cite{Georges1996,Metzner1989,Georges1992}. The method simplifies to the single-site DMFT for system size $N_c=1$, becomes exact for system size $N_c\rightarrow \infty$, and respects causal and conserving properties \cite{Hettler2000}. We refer the reader to the original literature \cite{Hettler1998} and a review \cite{Maier2005} for a detailed derivation but repeat some of the important aspects here.

DCA approximates the continuous lattice problem by an auxiliary discrete cluster problem with momenta $\tK, \tK', \tQ$ and modified cluster propagators $G^c(\omega_n,\tK)$, cluster self-energies $\Sigma^c(\omega_n, \tK)$, and cluster interactions $U+V^c(\tQ)$. The momentum points of the discrete cluster are embedded in the continuous Brillouin zone of the lattice, such that for  every lattice momentum $\tk$ we write $\tk=\tK+\zeta$, with $\tK$ a cluster momentum point close to $\tk$ (see left panel of Fig.~\ref{fig:DCA}). The regions over which the average is taken, the so-called `momentum patches', are centered around a cluster momentum $\tK$ and have equal size. The attribution of $k$-points to $\tK$ patches not unique~\cite{Gull2010,Ferrero2010,Staar2016}.

The DCA propagators are given by the `coarse-grained' patch averages
\begin{align}
G^c(\omega_n,\tK)=\frac{1}{N} \sum_\zeta^N [(i\omega_n + \mu)\mathbb{1}-F_{\tK+\zeta}-\Sigma^c(\omega_n, \tK)]^{-1} \label{eq:clustergf}
\end{align}
the non-local DCA interactions are either left invariant,
\begin{align}
V^{c,i}(\tQ)=V(\tQ),
    \label{eqn:lattice_V}
\end{align}
or obtained by coarse-graining the interaction over patches,

\begin{align}
V^{c,a}(\tQ)=\frac{1}{N}\sum_\zeta^N V(\tQ+\zeta).
\label{eqn:DCA_V}
\end{align}
Both $V^{c,i}(\tQ)$ and $V^{c,a}(\tQ)$ have been used in practice \cite{Arita2004,Wu2014,Terletska2018,Jiang2018}. The self-energies $\Sigma^c(\omega_n, \tK)$ are obtained by evaluating a diagrammatic skeleton series with propagators $G^c$ and interactions $U+V^{c,a/i}$ to infinite order, which can be achieved via the solution of a quantum impurity problem \cite{Georges1992,Zgid2017}. Because Eq.~\ref{eq:clustergf} implicitly depends on $\Sigma^c$, the quantum impurity problem needs to be solved self-consistently until convergence.

As the cluster size is increased, the area over which $\zeta$ is averaged decreases and the method converges to the solution of the infinite lattice system. As in the case of periodic boundary conditions, local quantities, such as the energy, converge quadratically in the linear extent of the system \cite{Maier2005}.

\subsubsection{Commonalities and differences between DCA and twist-averaging}\label{subsubsec:dca_twist}
Twist-averaging and the DCA technique appear remarkably similar when written in this form (Eqs.~\ref{eq:twistedgf}~and~\ref{eq:clustergf}). Both rely on a summation over shifted momenta $\zeta$ which accounts for strong variations in the one-body part of the Hamiltonian, and both exhibit the same quadratic convergence to the infinite system size limit.  The major difference is in the place of the average. In twist-averaging, every diagram is obtained for a definite twist angle $\zeta$. In the DCA, the averaging takes place inside the propagator (and, in the case of $V^{c,a}$, inside the interaction vertex). The system is then solved for all twist angles at once, but the summation over twist angles leads to an implicit dependence that requires the DCA to be solved self-consistently.

The implicit averaging over momentum patches in the Brillouin zone in Eq.~\ref{eq:clustergf} has a further consequence: momentum conservation is only guaranteed within momentum patches located around $\tK$, rather than for each scattering vertex individually. This is illustrated in the two middle panels of Fig.~\ref{fig:DCA}, where DCA diagrams violate momentum  conservation at each vertex within $\zeta, \zeta'$, and $\zeta''$. On the level of the averaged cluster quantities, the momentum conservation is restored (right panel of Fig.~\ref{fig:DCA}). Since the momentum conservation violation occurs on the scale of the grid spacing, the approximation will scale to zero at the same rate as the grid discretization errors as the system size is increased.

\subsection{Solution of the finite system}\label{subsec:solver}
The Hubbard model can be studied with numerically exact methods on moderate-to-large systems \cite{LeBlanc2015}. However, since this work predominantly discusses finite size methodologies, we limit this study to the fully self-consistent $GW$ method \cite{Hedin1965}. Results from GW are expected to show similar finite size behavior as results from numerically exact methods but, because of the low polynomial scaling \cite{Yeh2022}, allow access to the large system sizes needed to rigorously show finite size convergence. Where numerically exact results are available for finite size extrapolations \cite{Fuchs2011,Fuchs2011c,Gull2011,PhysRevB.105.045109}, we have validated our conclusions against these results.

The GW self-energy in the Matsubara    frequency domain is
    \begin{align}
        \Sigma_\sigma(\omega_n,\tk) &= \Sigma_\sigma^{\infty}(\tk) \nonumber \\ - \frac{1}{\beta N_k}& \sum_{\tq,m,\sigma'}
        G_{\sigma'}(\omega_n + \Omega_m, \tk+\tq)
        W(\Omega_m, \tq),
        \label{eqn:SigmaGW}
    \end{align}
    where $\beta$ is the inverse temperature, $\omega_n = (2n+1)\frac{\pi}{\beta}$ denotes fermionic Matsubara frequencies and
    $\Omega_n = 2n\frac{\pi}{\beta}$ bosonic Matsubara frequencies. $W(\Omega_m, \tq)$ is the dynamical screened interaction,
    $G(\omega_n, \tk)$ the Matsubara frequency Green's function, and the static part of the self-energy is
    \begin{align}
        \Sigma^{\infty}(\tk) = \frac{1}{N_k} \sum_{\tq,n} (U+V(\tq)) \rho_{\tk+\tq},
    \end{align}
    with $\rho_\tk = \frac{1}{\beta}\sum_n G(\omega_n,\tk)$ the single-particle density matrix.
    Starting from the non-interacting Green's function:
    \begin{align}
        G^{0}(\omega_n, \tk) = \left( i\omega_n + \mu - H^{0}_{\tk} \right)^{-1}
        \label{eqn:SigmaInf}
    \end{align}
    we solve Eq.~\ref{eqn:SigmaGW} self-consistently to obtain the GW approximation. This requires solving the
    Dyson equation
    \begin{align}
        G(\omega_n, \tk) = G^{0}(\omega_n, \tk) + G^{0}(\omega_n, \tk) \Sigma(\omega_n, \tk) G(\omega_n, \tk).
    \end{align}
For a detailed description of the GW formalism used in this work see Ref.~\cite{Yeh2022}.

 \begin{figure}[h!tb]
     \centering
     \includegraphics[width=0.47\columnwidth]{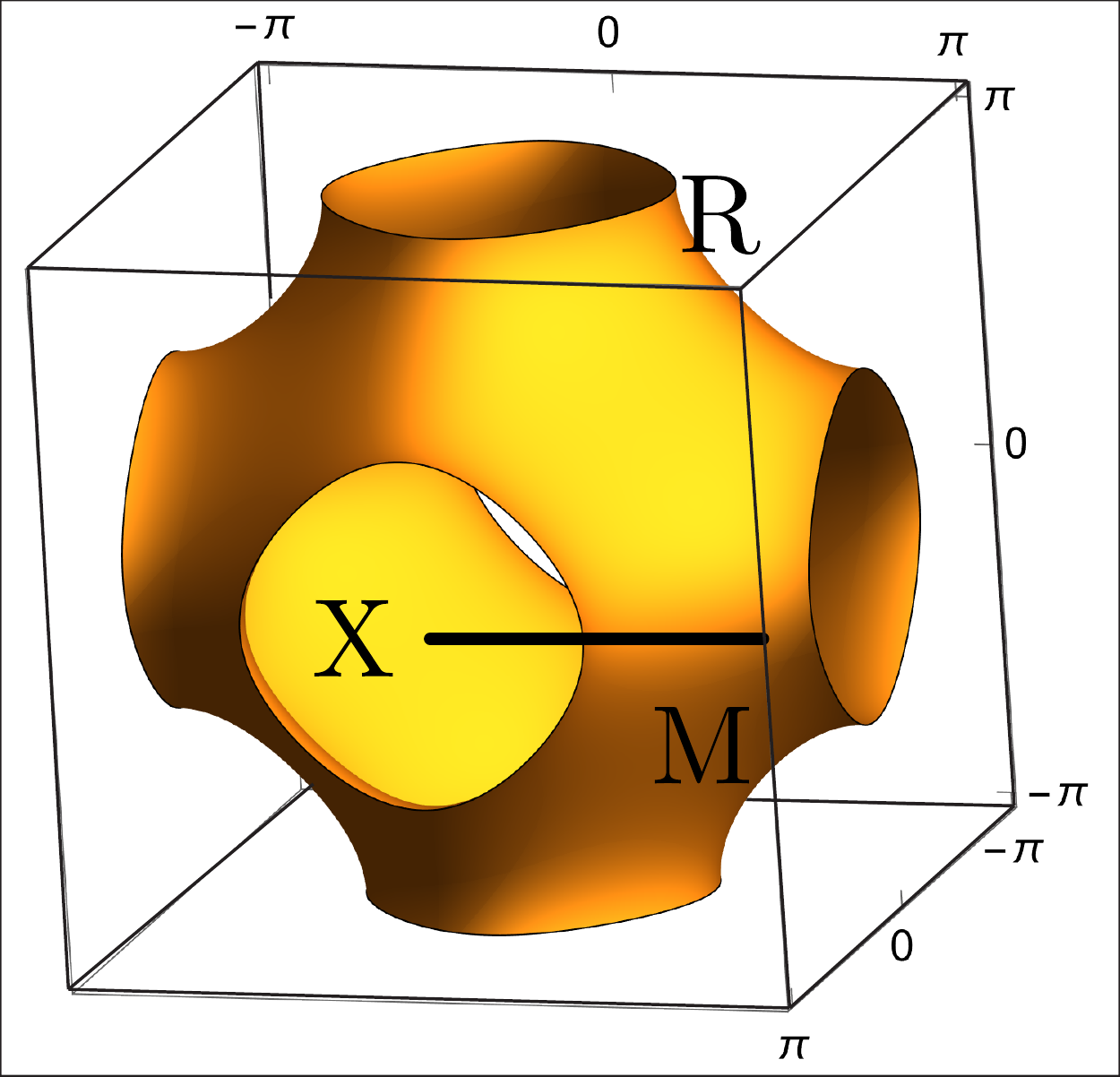}
     \includegraphics[width=0.47\columnwidth]{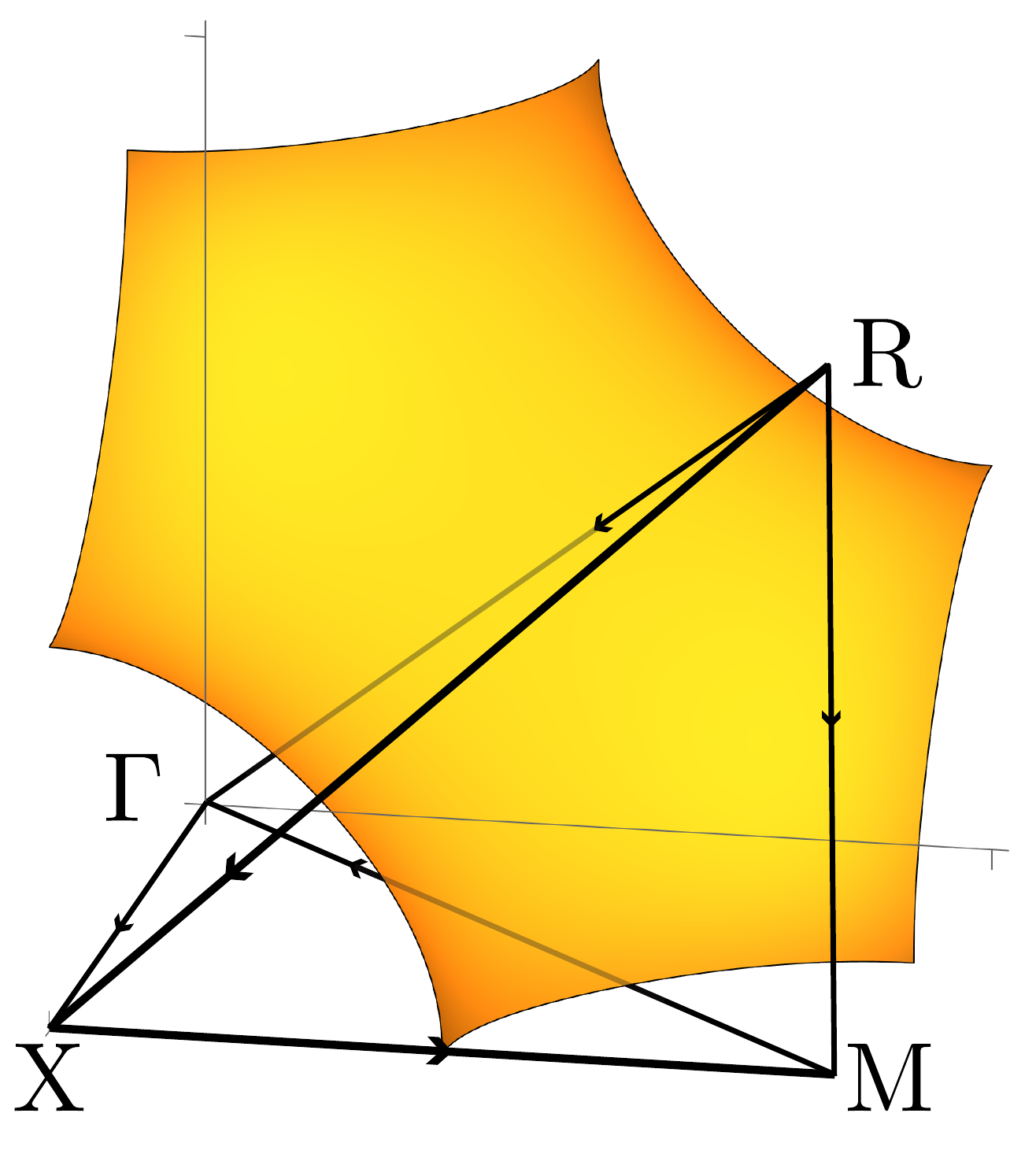}

     \includegraphics[width=0.9\columnwidth]{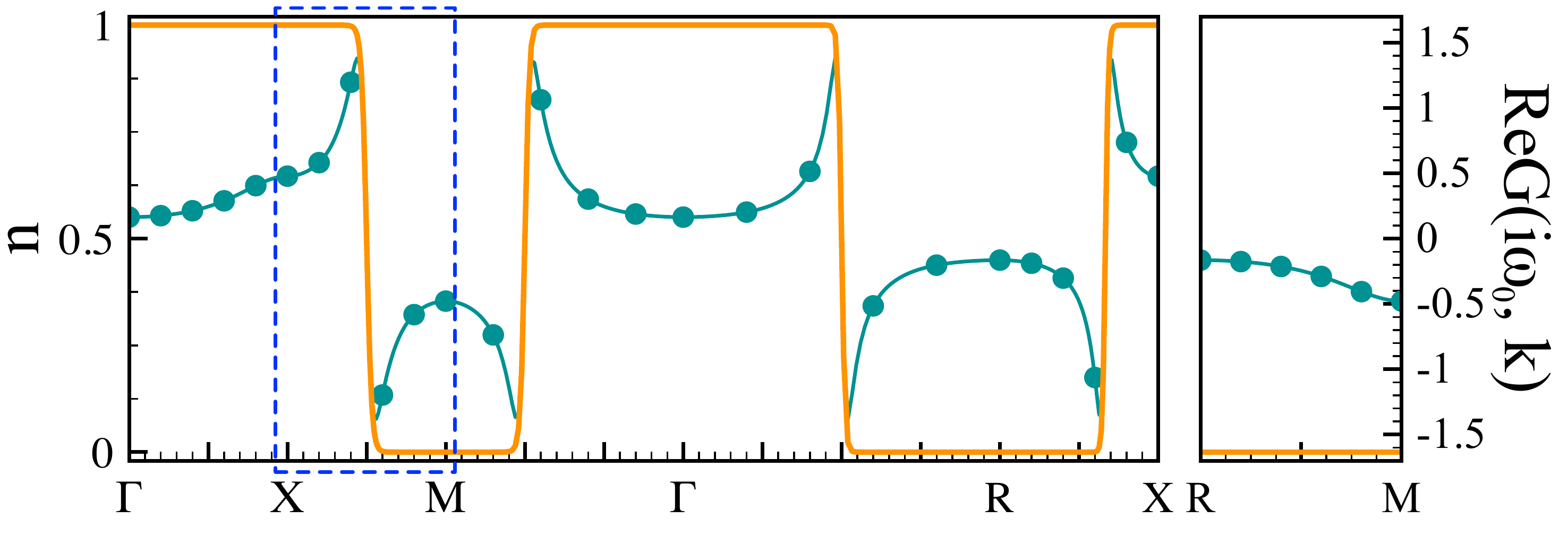}

     \caption{Top left: Fermi surface of the non-interacting 3D Hubbard model at half-filling in the simple cubic Brillouin zone, with special points $X$ and $M$ indicated.
     Top right: Special points and standard path $\Gamma X M \Gamma R X|R M$.
     Bottom: Density of the non-interacting model (orange, left axis) and lowest Matsubara frequency of the real part of the Green's
     function (blue, right axis) at $\beta=6.25$ and $U=4$ along the full high-symmetry path. For clarity the remainder of this paper shows momentum dependent
     quantities only along the path from $X$ to $M$, indicated by a blue square.}
     \label{fig:BZ}
 \end{figure}

 \begin{figure*}[tbh]
     \centering
     \includegraphics[width=0.9\columnwidth]{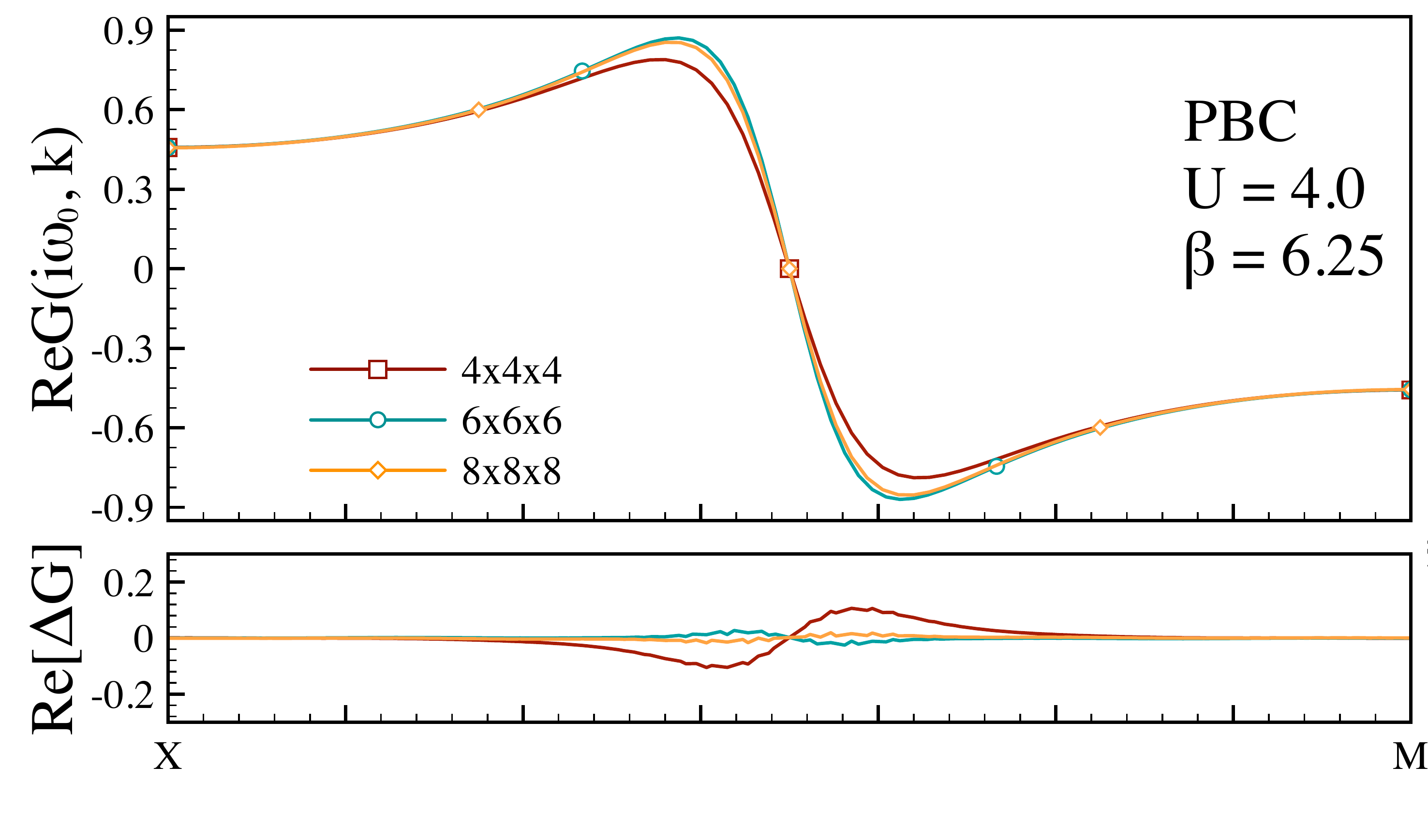}
     \includegraphics[width=0.9\columnwidth]{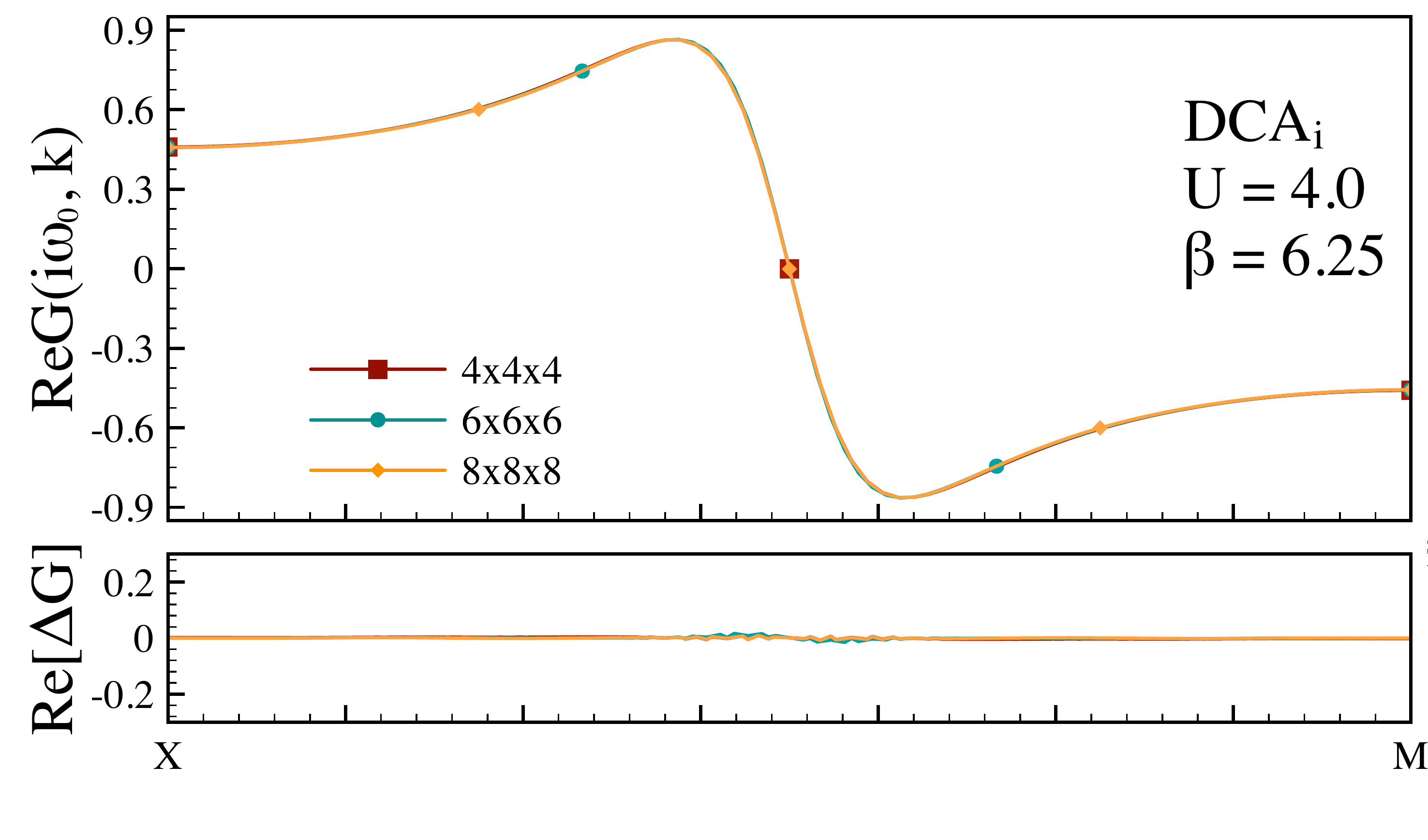}

     \includegraphics[width=0.9\columnwidth]{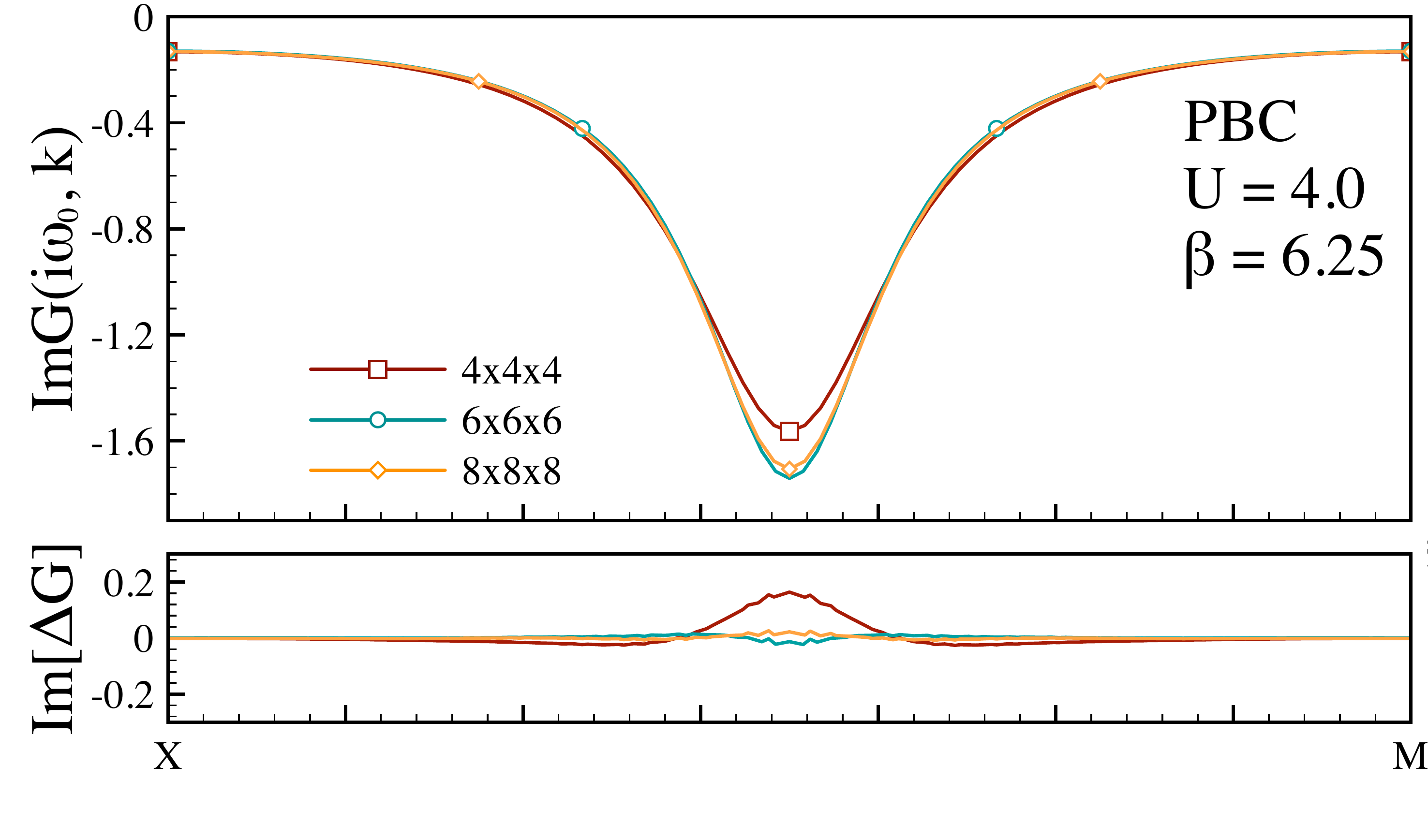}
     \includegraphics[width=0.9\columnwidth]{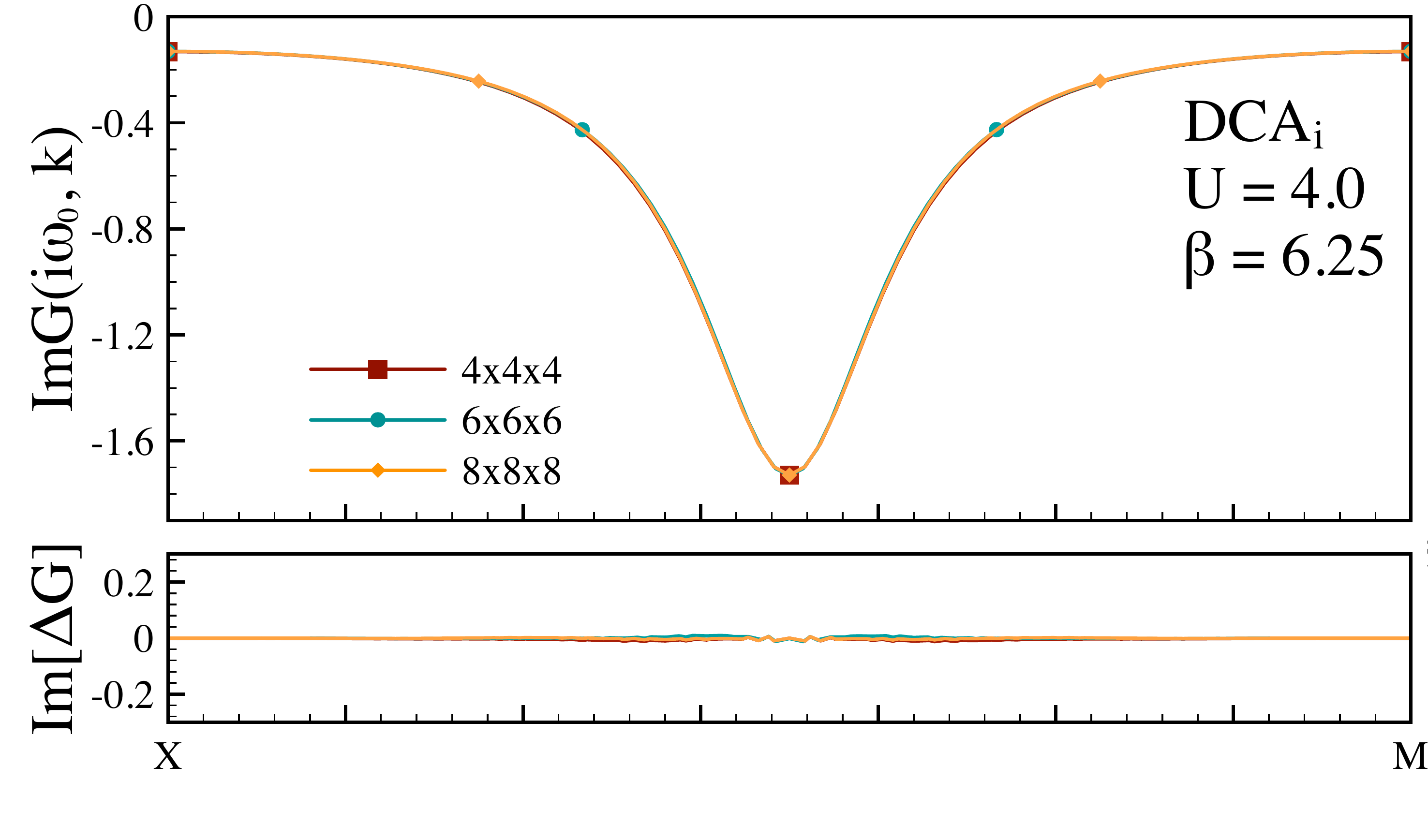}
     \caption{Momentum dependence of the real part of the Green's function at the lowest Matsubara frequency for the Hubbard
     model at $U=4$ and $\beta = 6.25$ for different system sizes.
     Left panels: Results for clusters with periodic boundary conditions.
     Right panels: DCA.
     Top row: Real part of the Green's function at lowest Matsubara frequency along a chosen region of the high symmetry path.
     Bottom row: Imaginary part of the Green's function at lowest Matsubara frequency along a chosen region of the high
     symmetry path.
     Bottom panels: Deviation from DCA results on a 12$\times$12$\times$12 lattice.
     Symbols correspond to actual data points. Smooth momentum dependence is obtained by Wannier interpolation.}
     \label{fig:Hub_G_U4}
 \end{figure*}

 \section{Results}\label{sec:results}
    We consider three different models in this paper: The 3D Hubbard model, the extended Hubbard model and the extended Hubbard
    model with Yukawa interactions. Each model is considered at a weak-to-intermediate interaction strength where 
    single-particle finite-size effects and correlation lengths are expected to be large (see Ref.~\cite{PhysRevB.105.045109} for the Hubbard case).

 Since much of this work discusses momentum-dependent quantities, we choose a standard way of plotting a representative part of the momentum dependence, rather than the full path in the Brillouin zone. The top left panel of Fig.~\ref{fig:BZ} shows the Fermi surface of the half-filled 3D Hubbard model. The top right panel illustrates the standard  high symmetry path in the irreducible wedge, along with a segment of the Brillouin zone, and the bottom panel shows the density and the lowest frequency of the real part of the Green's function along this standard path. Interesting physics with large finite size corrections occurs where the path intersects the Fermi surface, such as in the segment from $X$ to $M$. We therefore choose $X$ to $M$ as a representative path for display in the remainder of the paper (blue dashed section).

 \begin{figure*}[bth]
     \centering
     \includegraphics[width=0.9\columnwidth]{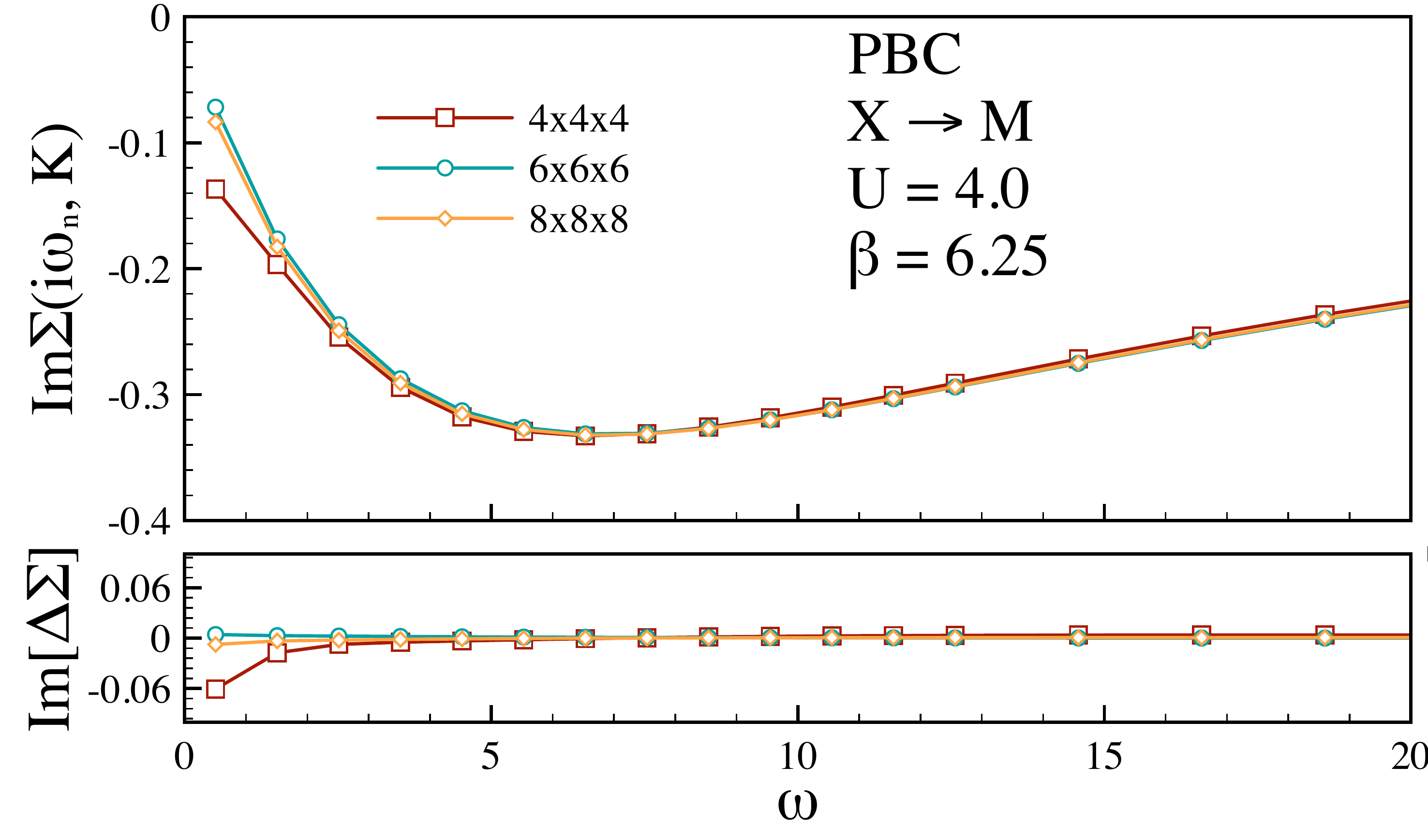}
     \includegraphics[width=0.9\columnwidth]{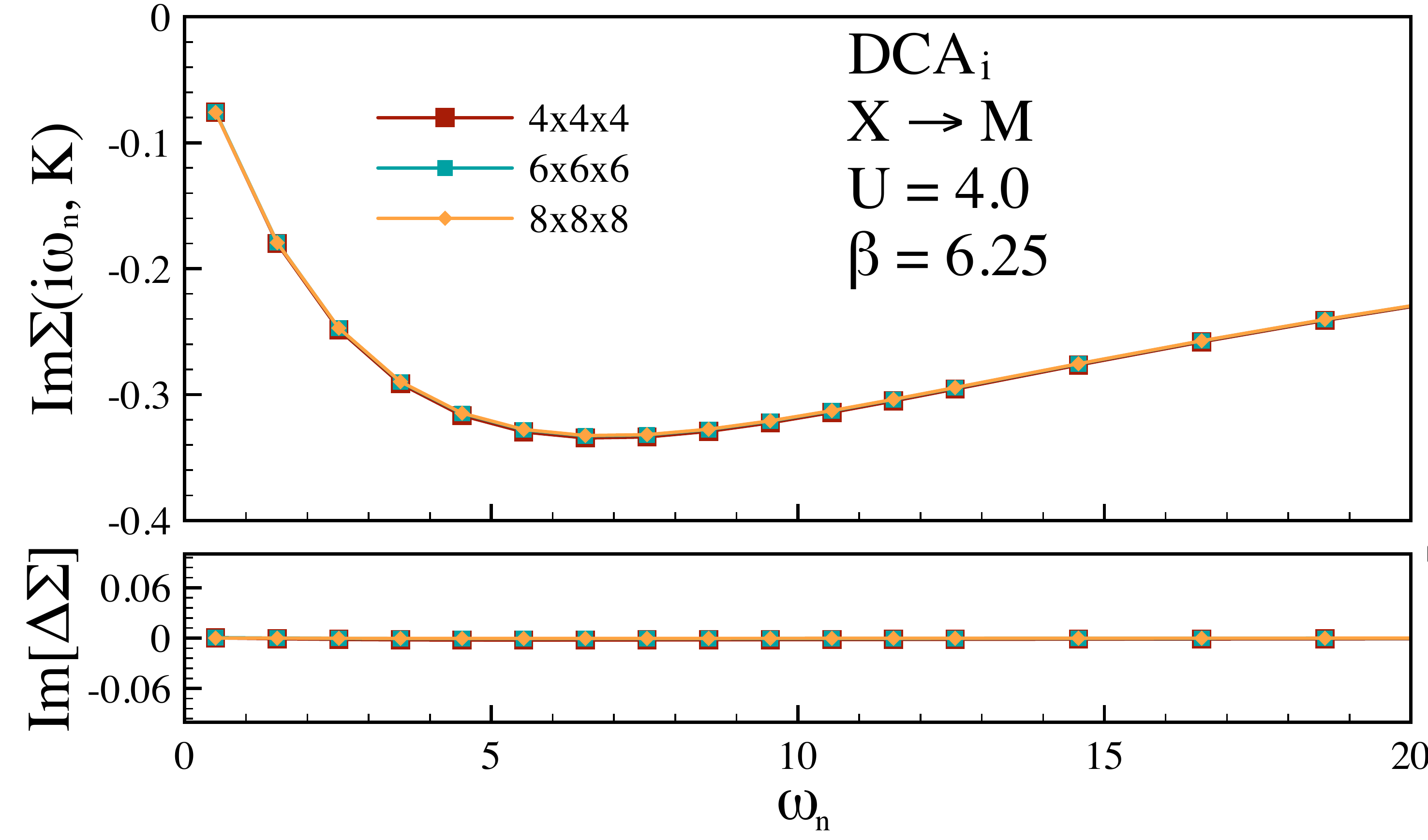}
     \caption{Imaginary part of the GW self-energy for the Hubbard model at $U=4.0$ and $\beta=6.25$ between the $X$ and
         $M$ points, for system sizes indicated.
         Left panels: Cluster with periodic boundary conditions.
         Right panels: DCA. Bottom insets show the deviation from DCA results on a 12$\times$12$\times$12 lattice.
     }
     \label{fig:Hub_SIGMA_U4}
 \end{figure*}

    \subsection{Local interaction: Hubbard model}\label{subsec:local}
We first analyze finite size effects in the half-filled Hubbard model in three dimensions.  Due to the locality of the interactions,
all finite-size effects originate from the one-body part of the Hamiltonian. Finite size effects consist of one-body  effects
 (or Brillouin zone approximation errors) and of errors due to the truncation of the non-local part of the frequency-dependent self-energy to the finite system.
 The Hartree contribution is an analytically known constant at particle-hole symmetry; the Fock contribution is identically zero.
 We also note that in addition to finite size errors there is a substantial (and often dominant) method-specific $GW$ approximation error in all the results
 we present in this section. As noted in Sec.~\ref{subsec:solver}, we do not discus this error here (see e.g. Refs.~\cite{PhysRevB.105.045109,LeBlanc2015}), focusing instead exclusively on finite size effects.

 As an example system, we choose a repulsive interaction strength of $U=4.0$ and set the overall hopping parameter to $t=1.0$.
 The temperature is $T=1/\beta=1/6.25$, which puts the system into the vicinity of the second-order phase transition for the GW method~\cite{PhysRevB.105.045109}.
 
 We briefly comment on the choice of interaction strength. For small $U$, correlations are typically long range but weak. For large $U$, correlations are often short-range but much stronger \cite{Qin2022}. Thus, to analyze finite size corrections, we choose an intermediate case of $U=4$ where finite size effects are long range but large enough in magnitude to have a noticeable effect.

The main panel of Fig.~\ref{fig:Hub_G_U4} shows the momentum
dependence of the Green's function at the lowest Matsubara frequency for systems of size 4$\times$4$\times$4, 6$\times$6$\times$6 and
8$\times$8$\times$8. The bottom panel additionally shows the difference to results obtained with DCA on a 12$\times$12$\times$12 lattice.
The top panels show the real parts, the bottom ones the imaginary parts.
Results for periodic boundary conditions (PBC) are shown in the left panels, those obtained with DCA in the right panels.
Data points $\tK$ are indicated with symbols; lines are generated from interpolated points obtained via Wannier interpolation.
PBC results are mostly converged by size $8\times 8\times 8$. As shown in the right panels, 
DCA eliminates  all system finite size effects for all the system sizes considered.

 \begin{figure*}[bt]
     \centering
     \includegraphics[width=0.9\columnwidth]{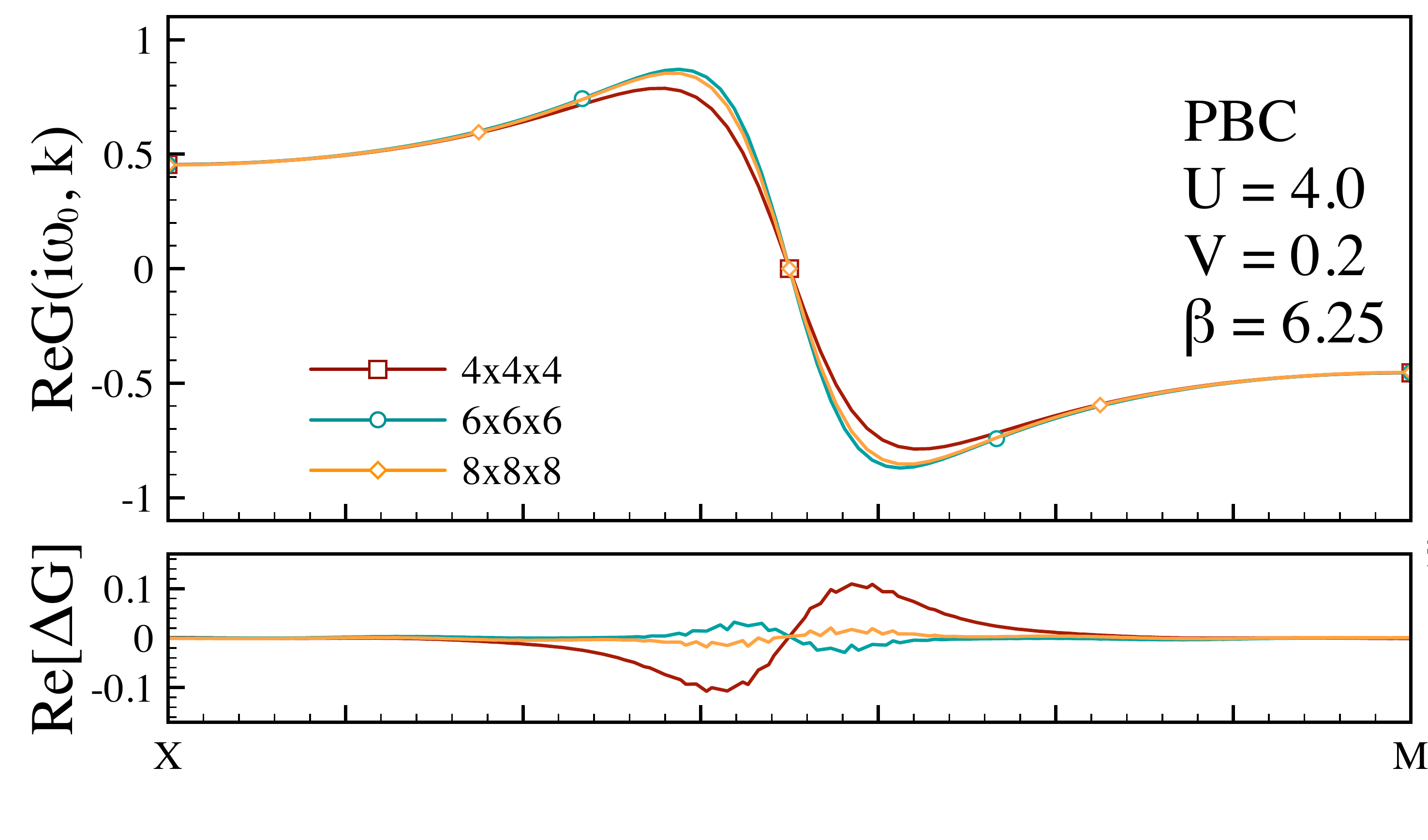}
     \includegraphics[width=0.9\columnwidth]{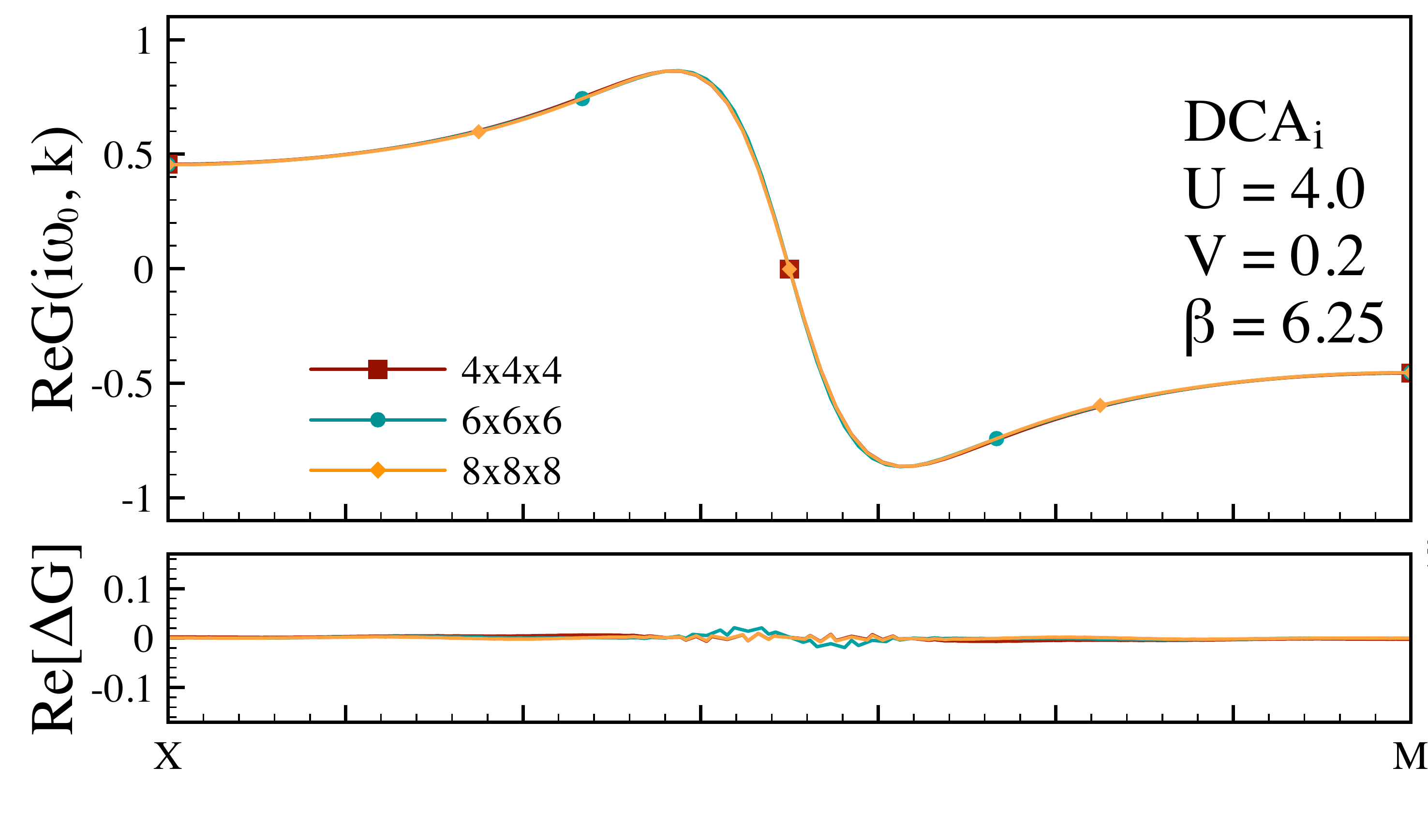}

     \includegraphics[width=0.9\columnwidth]{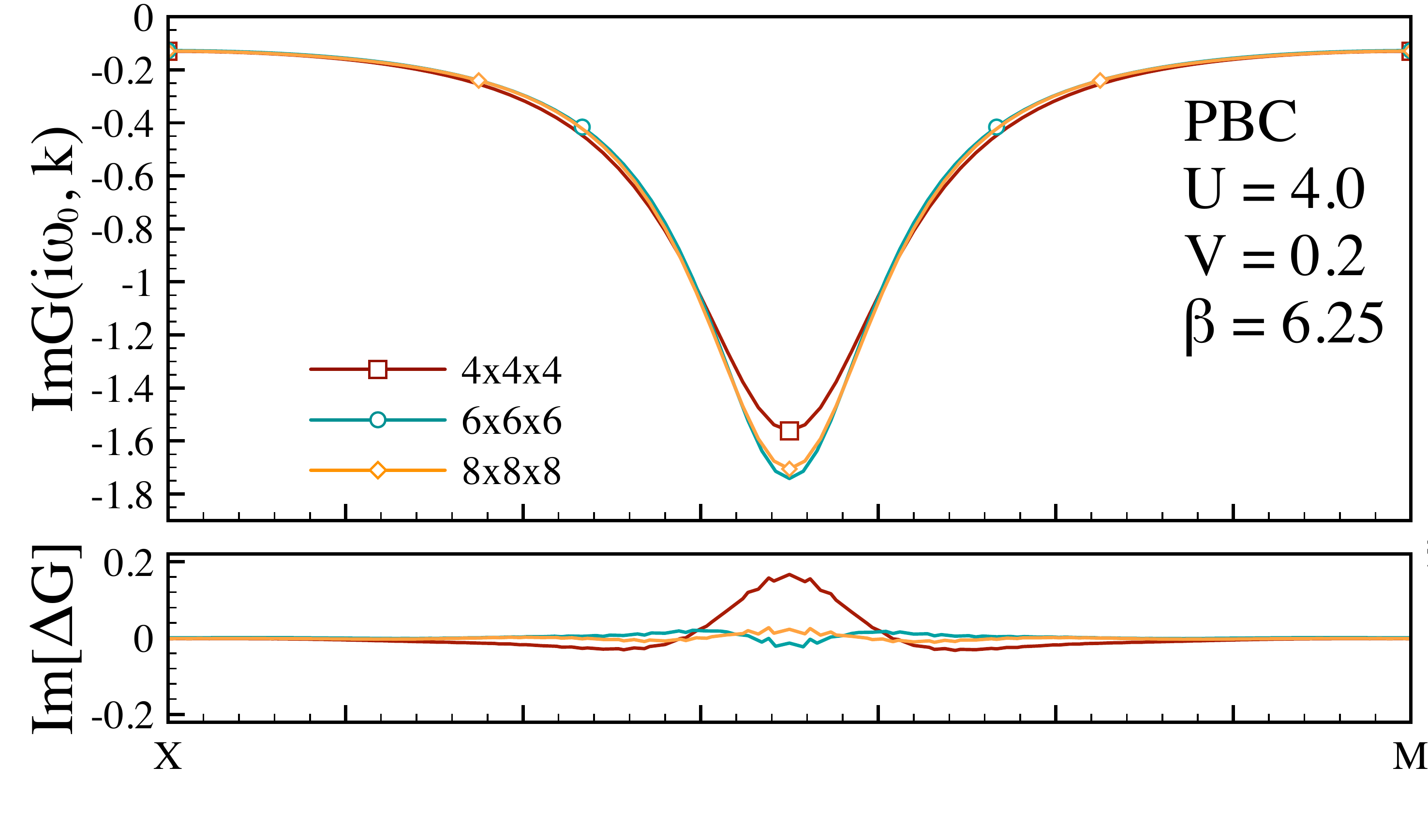}
     \includegraphics[width=0.9\columnwidth]{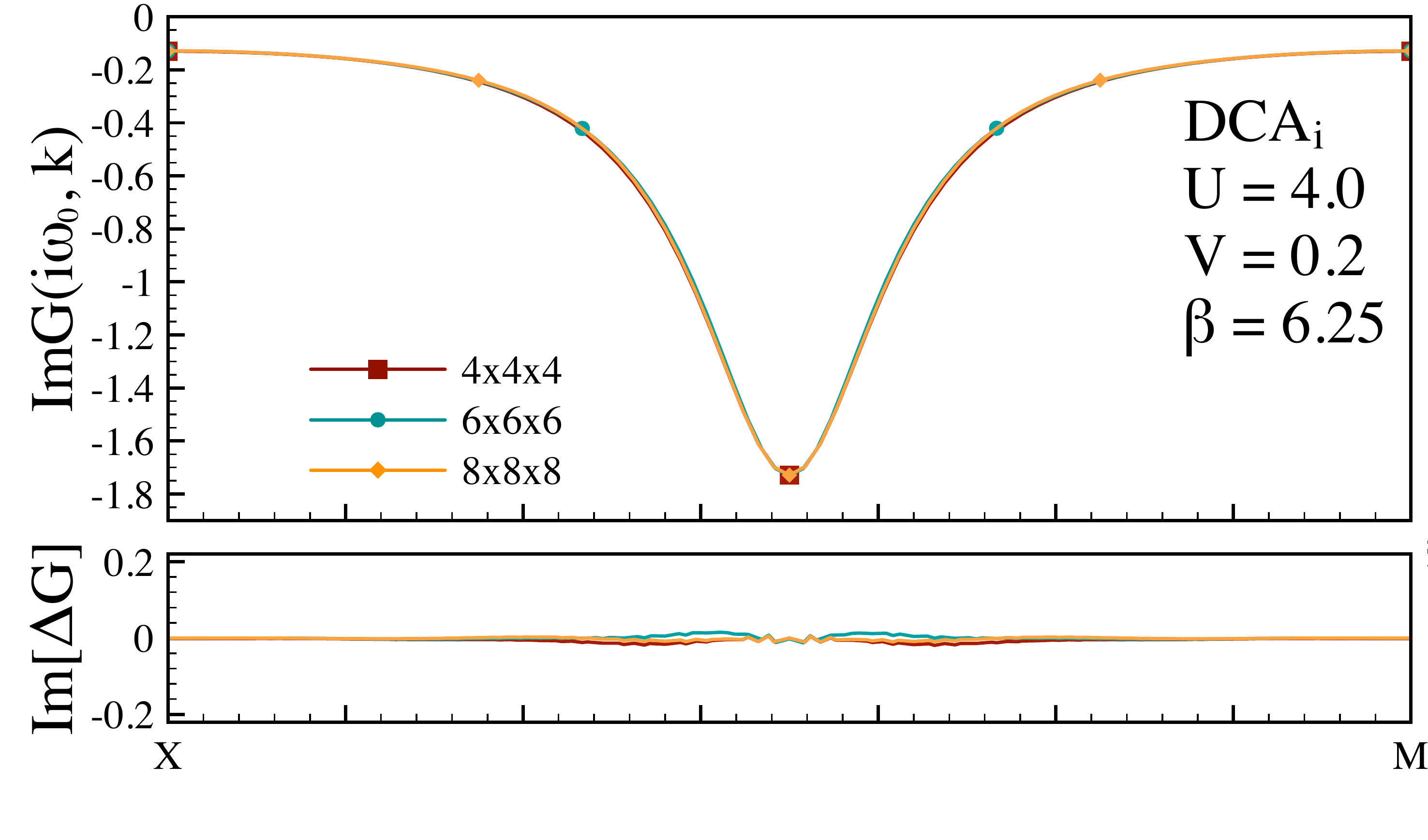}
     \caption{Momentum dependence of the real (top) and imaginary (bottom) part of the Green's function at the lowest Matsubara
     frequency for the extended Hubbard model at $U=4.0$, $V=0.2$ and $\beta=6.25$ in the GW approximation.
     For system sizes indicated.
     Left panels: Clusters with periodic boundary conditions.
     Right panels: DCA self-consistency.
     Bottom insets show the deviation from the DCA results on 12$\times$12$\times$12 lattice.
     Symbols correspond to actual data points. Smooth momentum dependence is obtained by Wannier interpolation.}
     \label{fig:G_U4}
 \end{figure*}

 \begin{figure*}[tb]
     \centering
     \includegraphics[width=0.9\columnwidth]{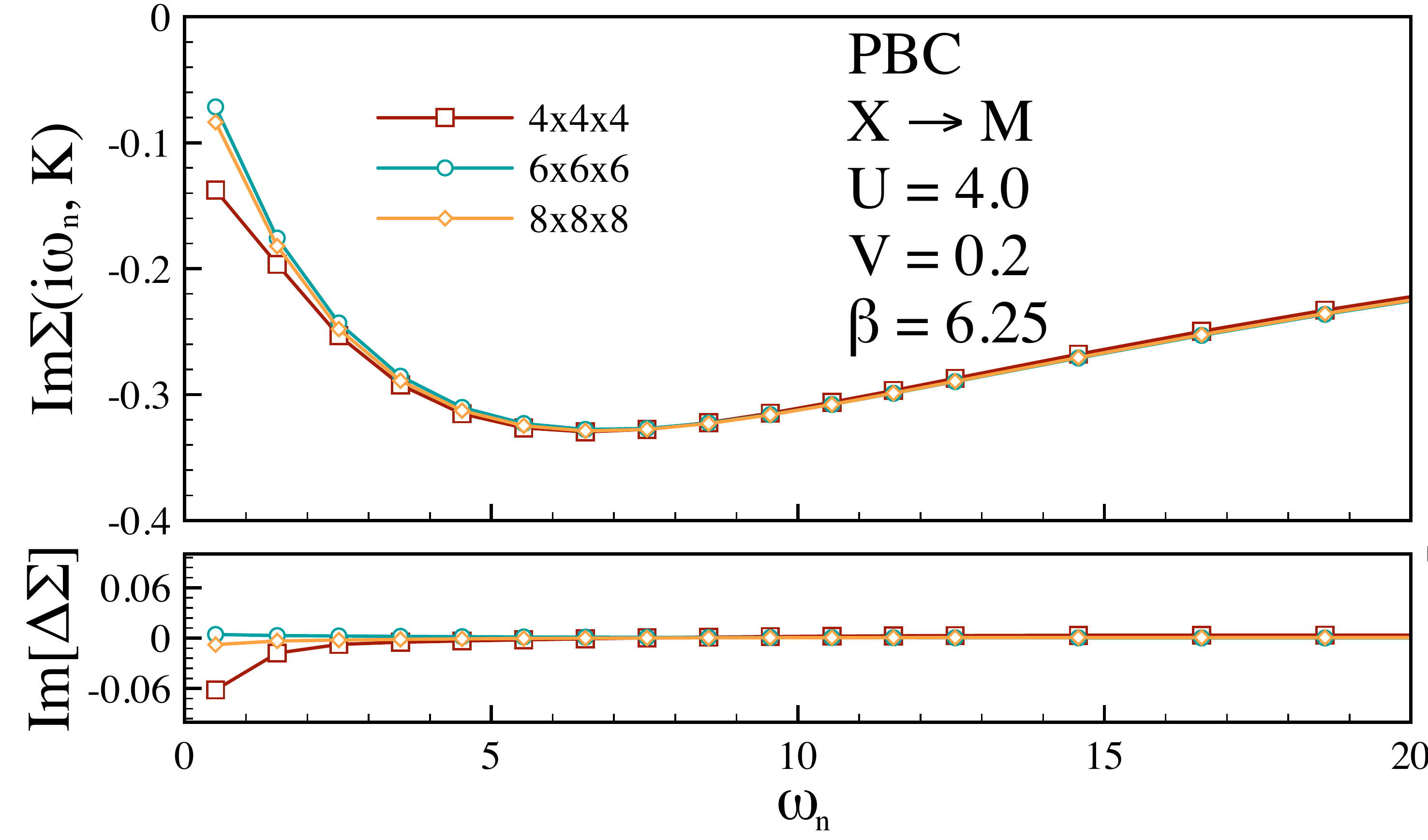}
     \includegraphics[width=0.9\columnwidth]{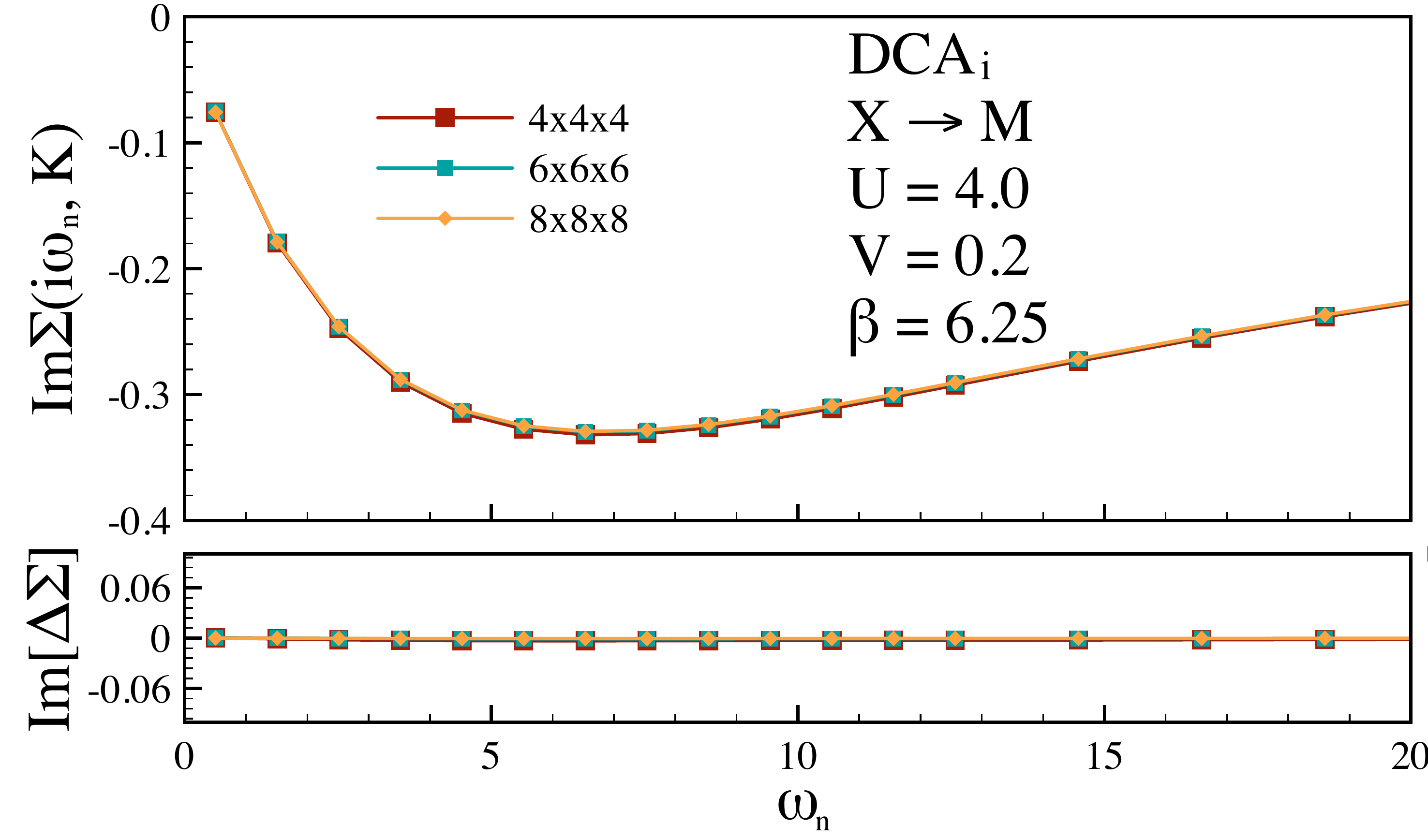}
     \caption{Imaginary part of the GW self-energy for the extended Hubbard model at $U=4.0$, $V=0.2$ and $\beta=6.25$ at the half-way point between
         $X$ and $M$, for system sizes indicated.
         Left panels: Clusters with periodic boundary conditions. Right panels: DCA.
         Bottom insets show the deviation from DCA results on a 12$\times$12$\times$12 lattice.
     }
     \label{fig:SIGMA_U4}
 \end{figure*}

 Fig.~\ref{fig:Hub_SIGMA_U4} shows the finite size convergence of the frequency-dependence of the imaginary part of the self-energy. We choose a point halfway between $X$ and $M$, lying on the non-interacting Fermi surface. As the high-frequency limit for the self-energy converges to an analytically known \cite{Rusakov2014} constant, all finite size corrections occur at low frequency. At the lowest Matsubara frequency, PBC self-energies show  deviations on systems of size $4\times4\times4$ but converge as the system size is
 increased towards $8\times$ 8$\times 8$. 
 DCA results, due to complete elimination of the BZ discretization error, converge for much smaller system size. 

The almost complete absence of finite size errors in DCA  can be understood as follows. In the regime studied here, correlations are weak and almost all finite-size effects in the GW
 solution of Hubbard model are due to the single-particle finite size effects or discretization of $H_0$. The DCA average over single-particle propagators eliminates these finite-size effects
 with its summation over all momenta. Non-local self-energy contributions have a shorter range than the smallest system size and do not result in visible finite size effects. On the other hand, PBC results visibly suffer from single-particle finite size effects.

 \subsection{Nearest-neighbor interaction: extended Hubbard model}\label{subsec:nn}
We next consider the extended Hubbard model at intermediate local interaction strength $U=6.0$ and nearest neighbor interaction $V_{12}=0.2$. Fig.~\ref{fig:G_U4} shows the momentum dependence of the Green's function at the lowest Matsubara frequency. 
Both PBC (left panel) and DCA (right panel) show only small finite-size effects at the fully occupied ($X$ point) and empty ($M$ point) regions of the Brillouin zone. However, close to the partially
 occupied momentum points, where correlation effects are expected to be most pronounced, PBC solutions exhibit a much slower
 finite size convergence in comparison to DCA. From the analysis of the Matsubara self-energy for a half-filled $\tK$-point
 (Fig.~\ref{fig:SIGMA_U4}) we see that finite size errors are mostly restricted to the low-frequency behavior of the dynamical self-energy.

\subsubsection*{Energetics and interaction coarse-graining}
Finite size approximations to the Hamiltonian result in approximation errors in the self-energy and the Green's function, which propagate to finite size errors in the energy. We calculate the correlation energy $\text{Tr} \Sigma G$, with $\Sigma$ the dynamical part of the self-energy,  and show relative errors $\Delta$ in comparison to the DCA result $E_\text{DCA}^{L=12}$ on a 12$\times$12$\times$12 lattice, which we believe to be our most accurate result.
Table~\ref{tab:energy_U4} shows $\Delta_\text{PBC}$ for naive PBC, $\Delta_{\text{TABC}_5}$ for twist-averaging with $5\times 5 \times 5=125$ and $\Delta_{\text{TABC}_7}$ for $7\times 7 \times 7=343$ twists, and for the two variants of DCA discussed in Sec.~\ref{subsubsec:dca}: non-coarse-grained interactions analogous to those used in the PBC and TABC methods, Eq.~\ref{eqn:lattice_V} ($\Delta_{\text{DCA}_i}$), and coarse-grained interactions according to  Eq.~\ref{eqn:DCA_V} ($\Delta_{\text{DCA}_a}$).

\begin{table}[bth]
        \begin{tabular}{p{0.1\columnwidth}|p{0.15\columnwidth}p{0.15\columnwidth}p{0.15\columnwidth}p{0.15\columnwidth}p{0.15\columnwidth}}
            L$_c$ & $\Delta_\text{PBC}$ & $\Delta_{\text{TABC}_5} $ & $\Delta_{\text{TABC}_7} $  & $\Delta_{\text{DCA}_i}$ & $\Delta_{\text{DCA}_a}$ \\
            \hline
            2 & 0.1796 & 0.1897 & 0.0810 & 0.0120 & 0.0106   \\
            4 & 0.0614 & 0.0016 & 0.0067 & 0.0007 & 0.0007   \\
            6 & 0.0040 & 0.0027 & 0.0017 & 0.0002 & 0.0002 \\
            8 & 0.0100 & 0.0004 & 0.0011 & 0.0001 & 0.0001
        \end{tabular}
        \caption{Relative correlation energy error, $\Delta E = |E - E^{DCA}_{Nc=12}|/E^{DCA}_{Nc=12}$, for the extended Hubbard model at $U=4$, $V=0.2$ and $\beta=6.25$ for
        different methods and system sizes. }
        \label{tab:energy_U4}
\end{table}

All finite size correction methods converge to the same result as a function of system size. Twist-averaging is clearly beneficial and leads to errors that are substantially smaller than those obtained with PBC on all but the smallest system. These results clearly illustrate the power of twist-averaging for energy calculations.

Little difference is evident between the two DCA methods, {i.e.} interaction coarse-graining and no interacting coarse-graining. This is consistent with previous results for the extended Hubbard model \cite{Wu2014,Terletska2018,Jiang2018} and generally expected for interactions that have little momentum dependence.

In addition to twist-avering over $5\times5\times5$ twists, we performed simulations for $3\times3\times3$ (not shown) and $7\times7\times7$ twists. While differences between $3\times3\times3$ and $5\times5\times5$ twists are substantial, results only marginally change  between $5\times5\times5$ and $7\times7\times7$ twists.

 \begin{figure*}[bt]
     \centering
     \includegraphics[width=0.9\columnwidth]{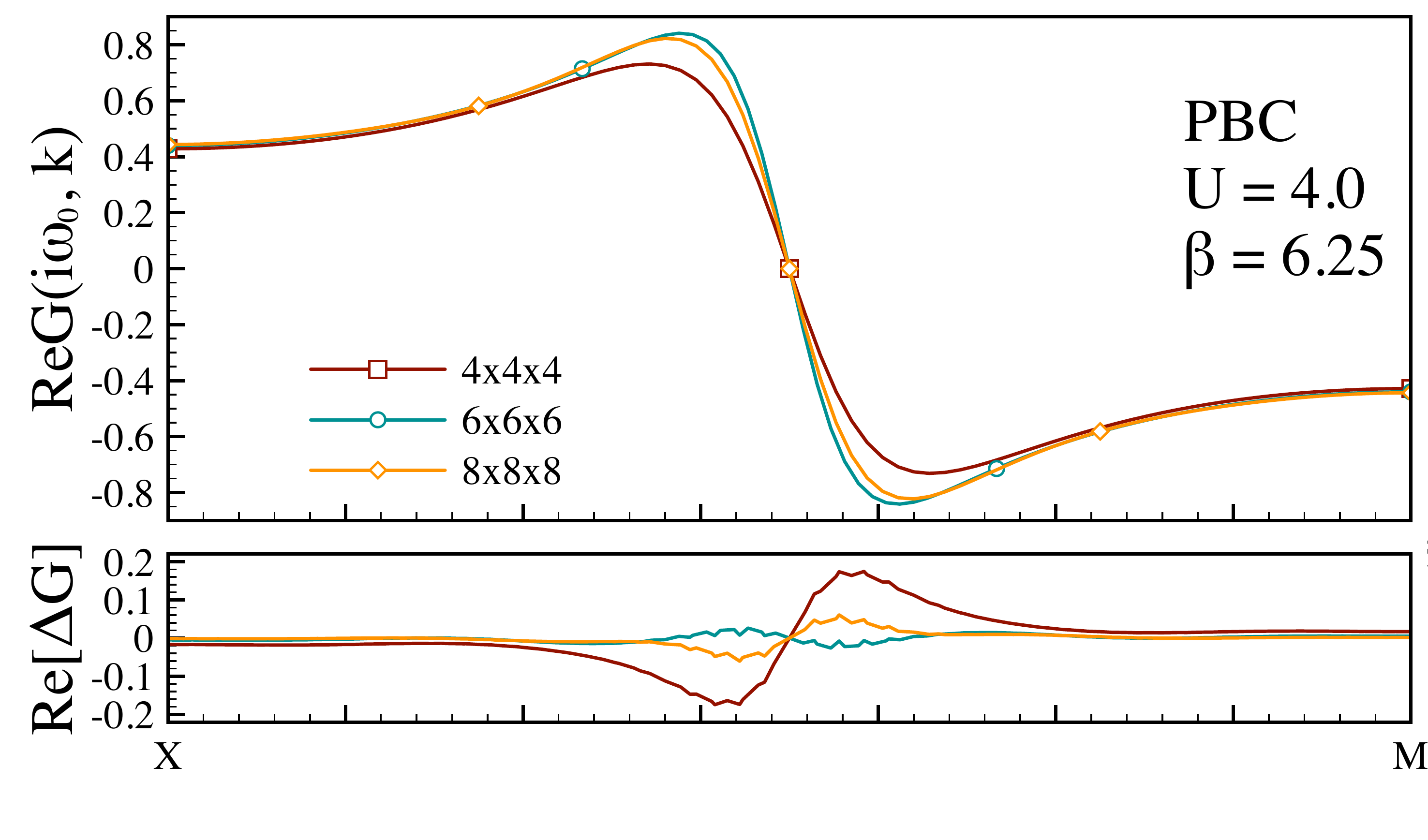}
     \includegraphics[width=0.9\columnwidth]{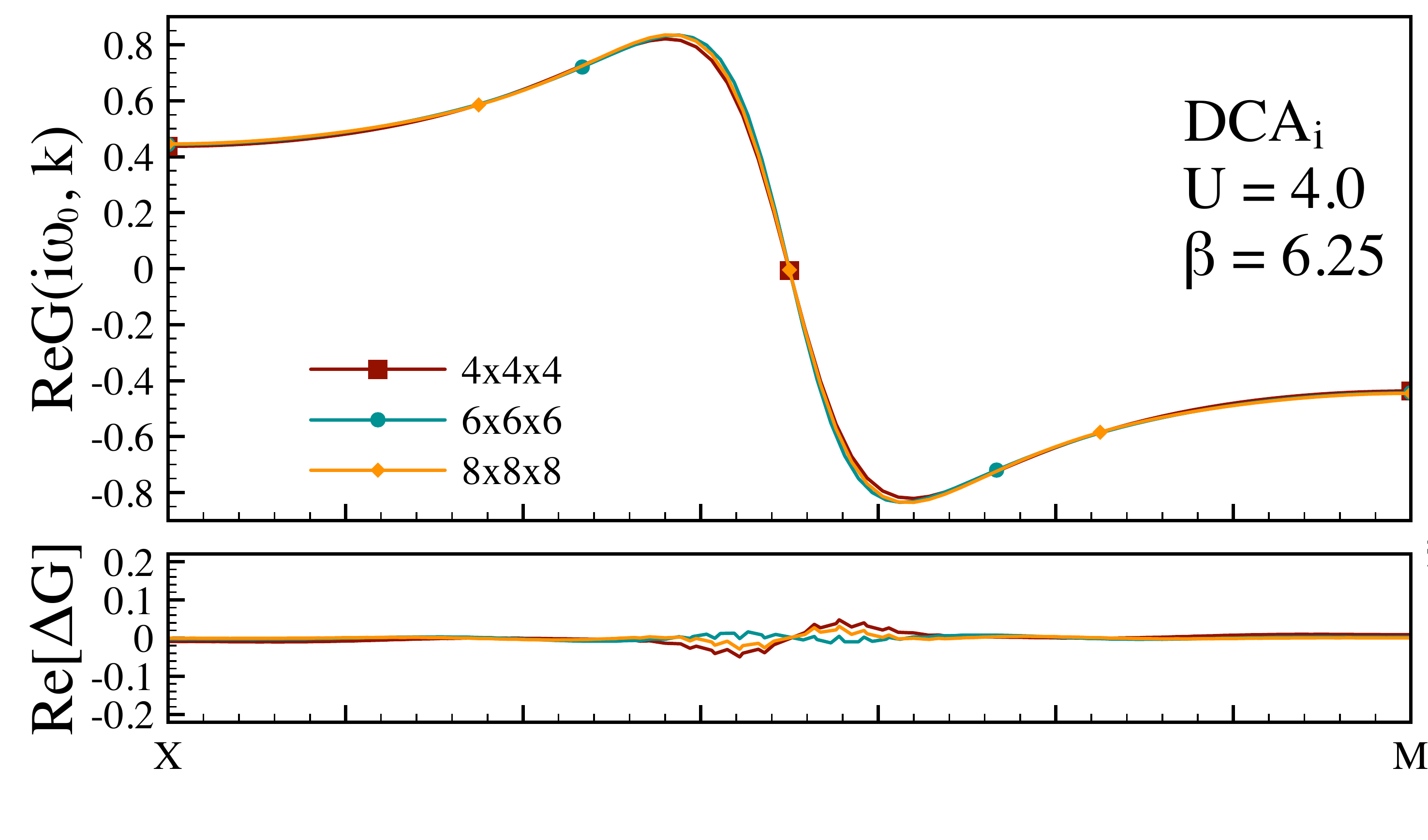}

     \includegraphics[width=0.9\columnwidth]{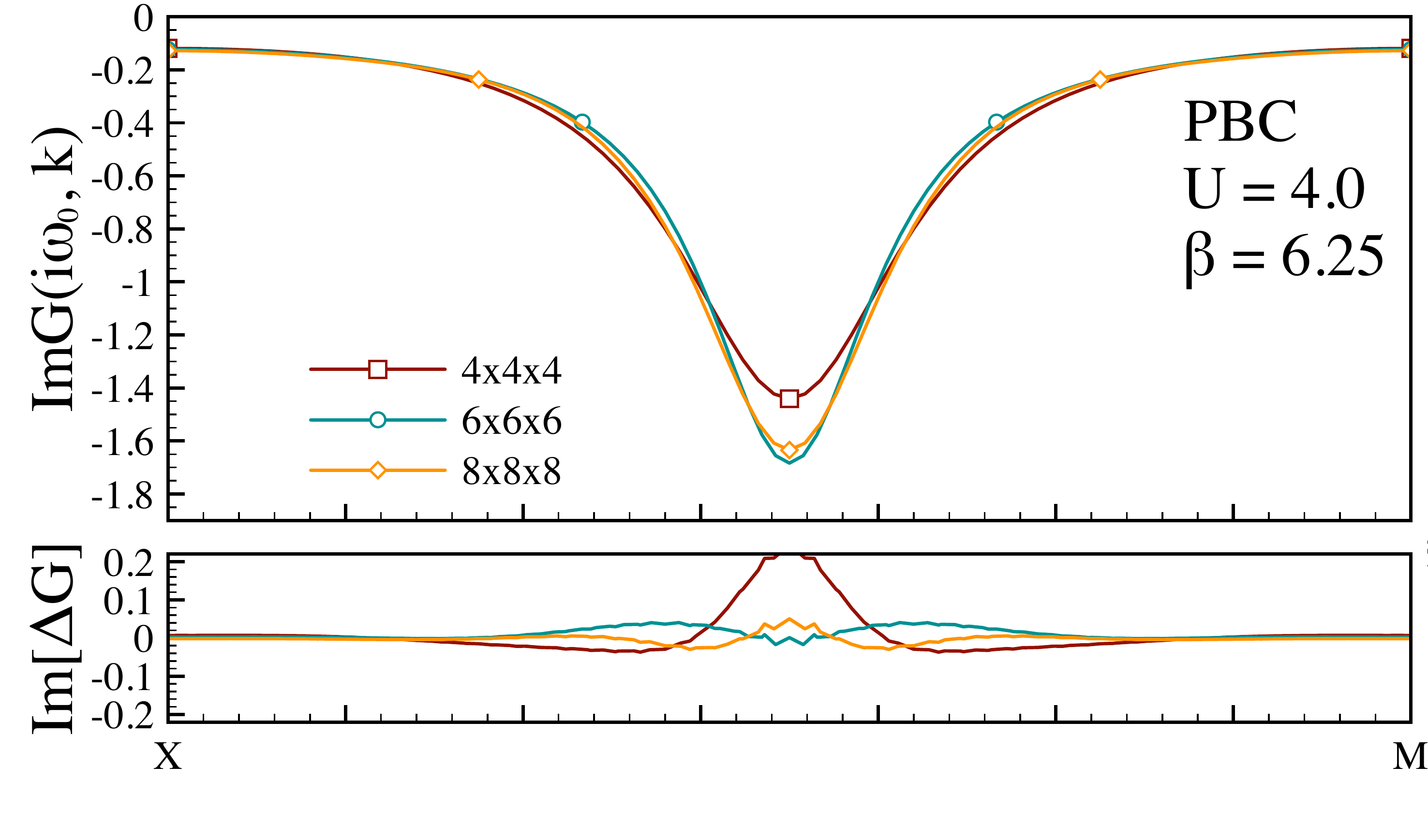}
     \includegraphics[width=0.9\columnwidth]{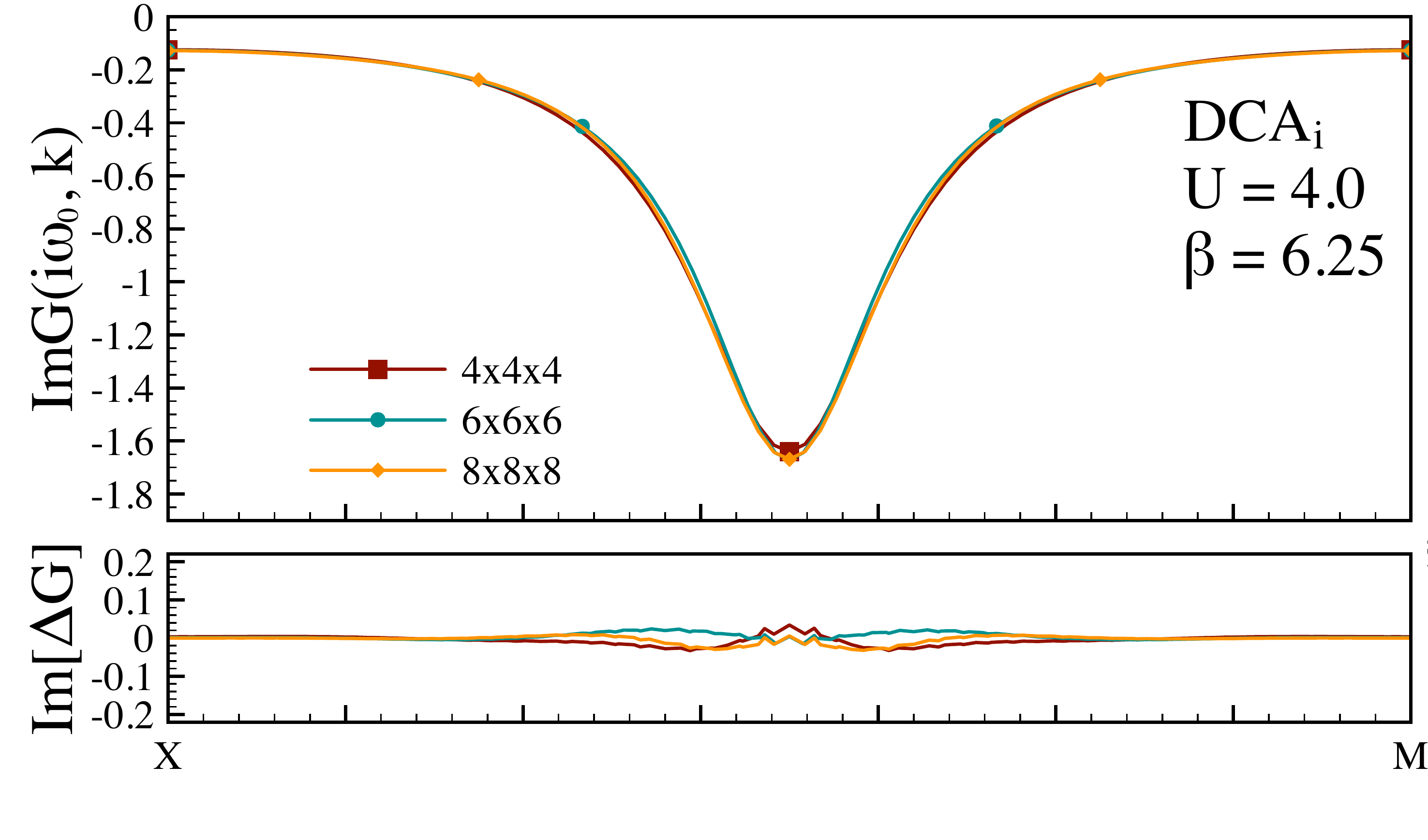}

     \caption{Momentum dependence of the real (top) and imaginary (bottom) part of the Green's function at lowest Matsubara
     frequency for the extended Hubbard model with Yukawa interaction at $U=4.0$, $V_0=0.2$, $\alpha=0.25$ and $\beta=6.25$,
         for system sizes indicated.
         Left panels: Clusters with periodic boundary conditions.
         Right panels: DCA self-consistency.
         Bottom insets show the deviation from DCA results on a 12$\times$12$\times$12 lattice.
         Symbols correspond to actual data points. Smooth momentum dependence is obtained by Wannier interpolation.}
     \label{fig:YUKAWA_G_U4}
 \end{figure*}

\subsection{Longer-range interaction: Yukawa Hubbard model}\label{subsec:long}
The Yukawa-Hubbard model is chosen to mimic long-range interactions without a divergent $1/r$ contribution.
Unlike in the extended Hubbard model, the interaction range may be larger than the finite size system considered.

To construct a model with interactions beyond the system size, 
we choose $\alpha=0.25$ in Eq.~\ref{eq:Yukawa}. In this case, the interactions
with the nearest neighbor outside a 4$\times$4$\times$4 cluster is approximately $10\%$ of the nearest-neighbor interaction.
We then normalize the strength $V_0$ of the Yukawa potential to obtain a nearest-neighbor interaction $V_{0} = 0.2$, comparable to one
chosen for extended Hubbard model. 

 \begin{figure*}[tb]
     \centering
     \includegraphics[width=0.45\textwidth]{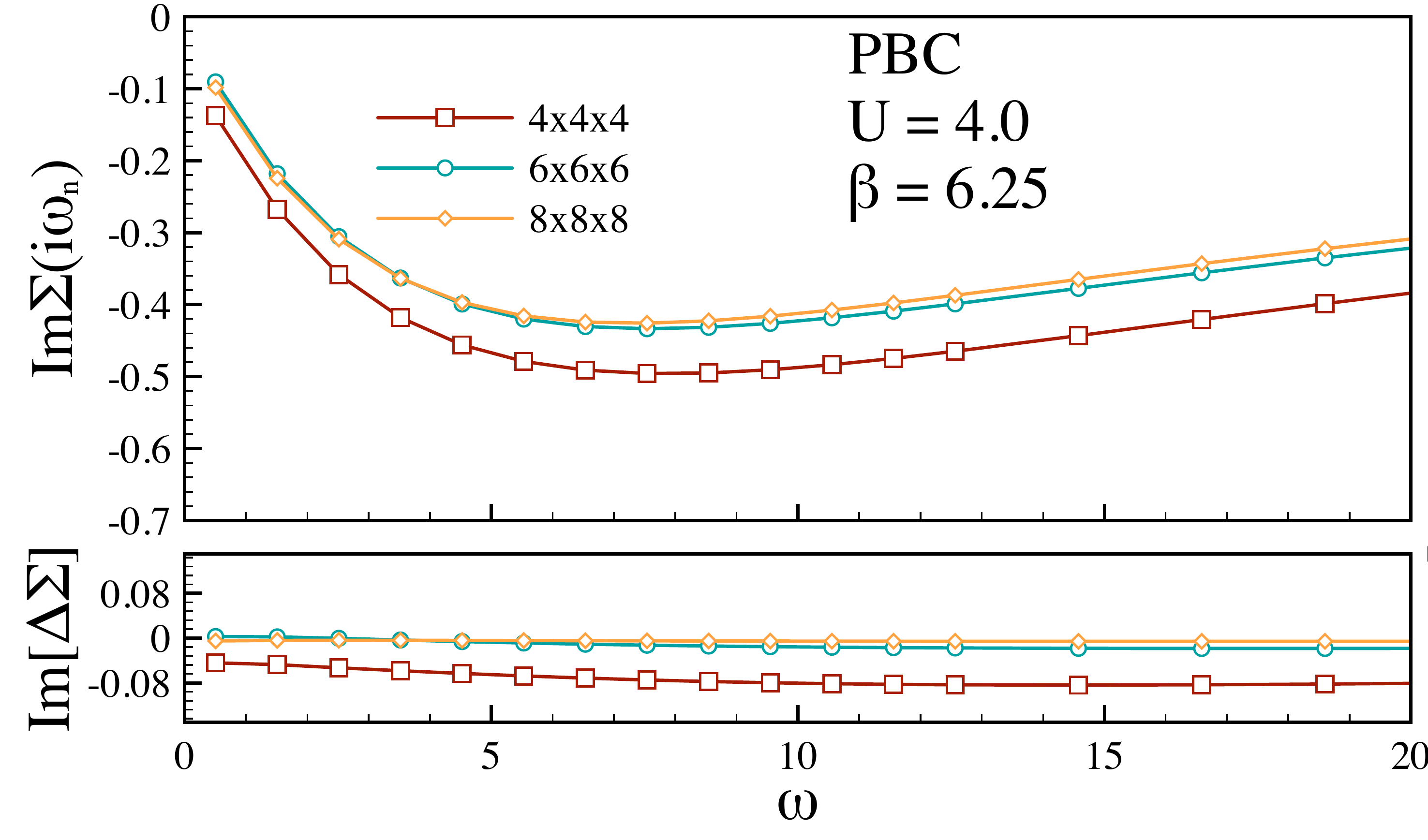}
     \includegraphics[width=0.45\textwidth]{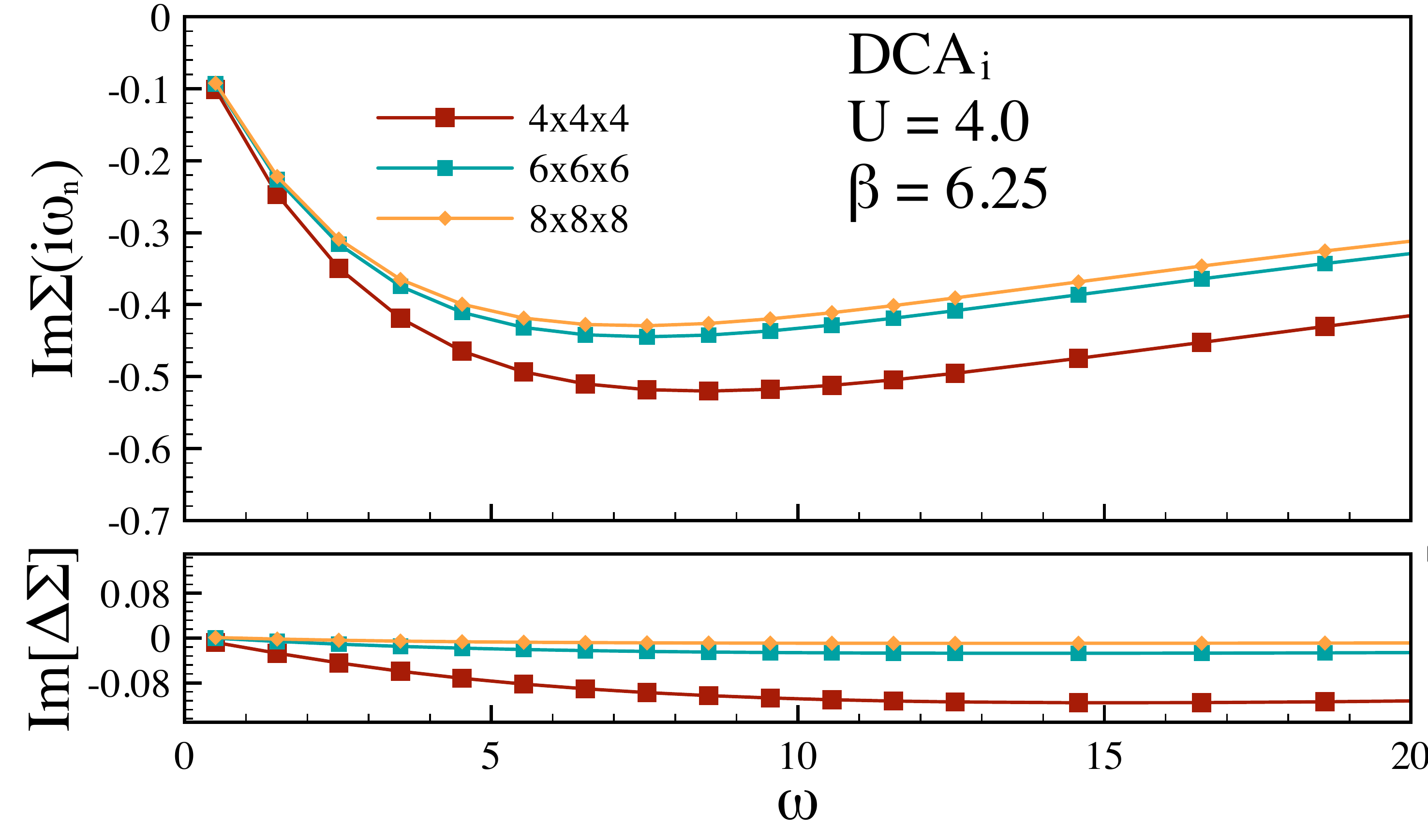}\\
     \includegraphics[width=0.45\textwidth]{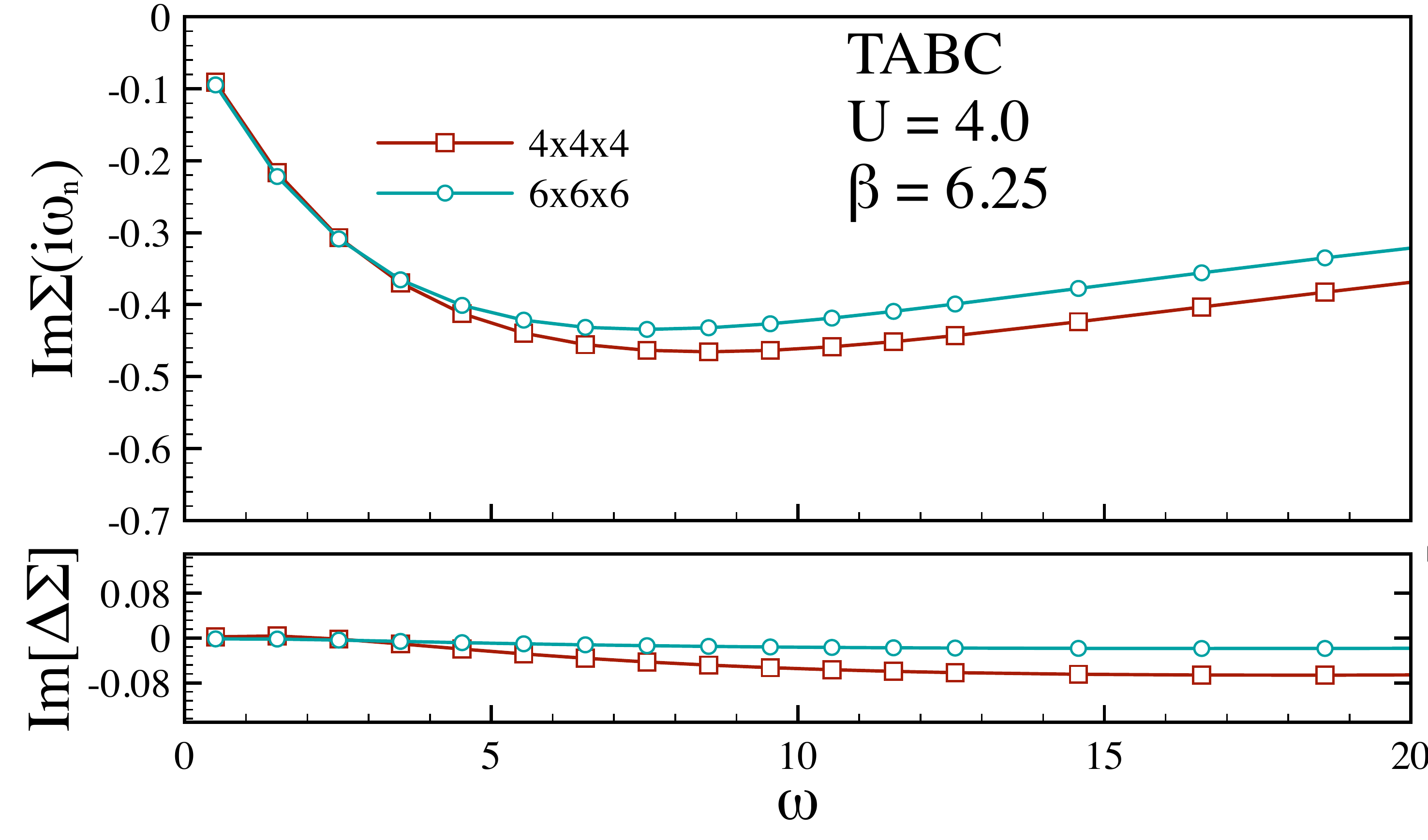}
     \includegraphics[width=0.45\textwidth]{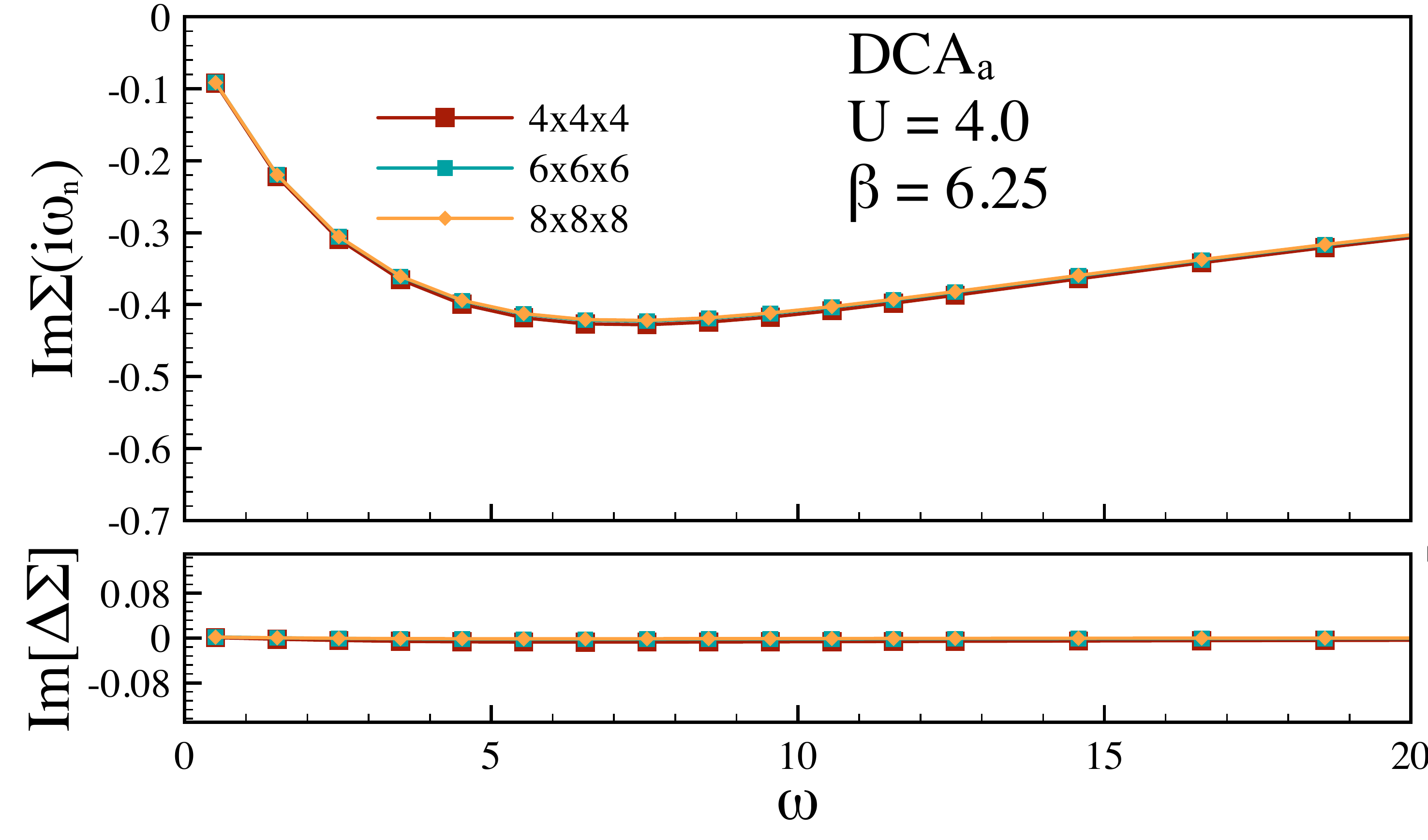}
     \caption{
     Frequency dependence of the imaginary part of the local self-energy for the extended Hubbard model with Yukawa interaction at $U=4.0$, $V_0=0.2$,
         $\alpha=0.25$ and $\beta=6.25$, for system sizes indicated. Top left: PBC. Rop right: DCA with invariant interactions. Bottom left: TABC (note that the $8\times8\times8$ system with 125 twists is beyond our capabilities). Bottom right: DCA with interaction coarse-graining. Bottom insets show the deviation from DCA
         results on a 12$\times$12$\times$12 lattice.
     }
     \label{fig:YU_SIGMA_U4}
 \end{figure*}

As above, we consider the weak-to-intermediate interaction strength $U=4.0$ and $\beta=6.25$. Fig.~\ref{fig:YUKAWA_G_U4} shows results 
for the Green's function  at the lowest Matsubara frequency. The overall behavior is similar to the one in Fig.~\ref{fig:G_U4} for the extended Hubbard model.
DCA results converge quickly at low frequency.

 \begin{figure}[tb]
     \centering
     \includegraphics[width=0.9\columnwidth]{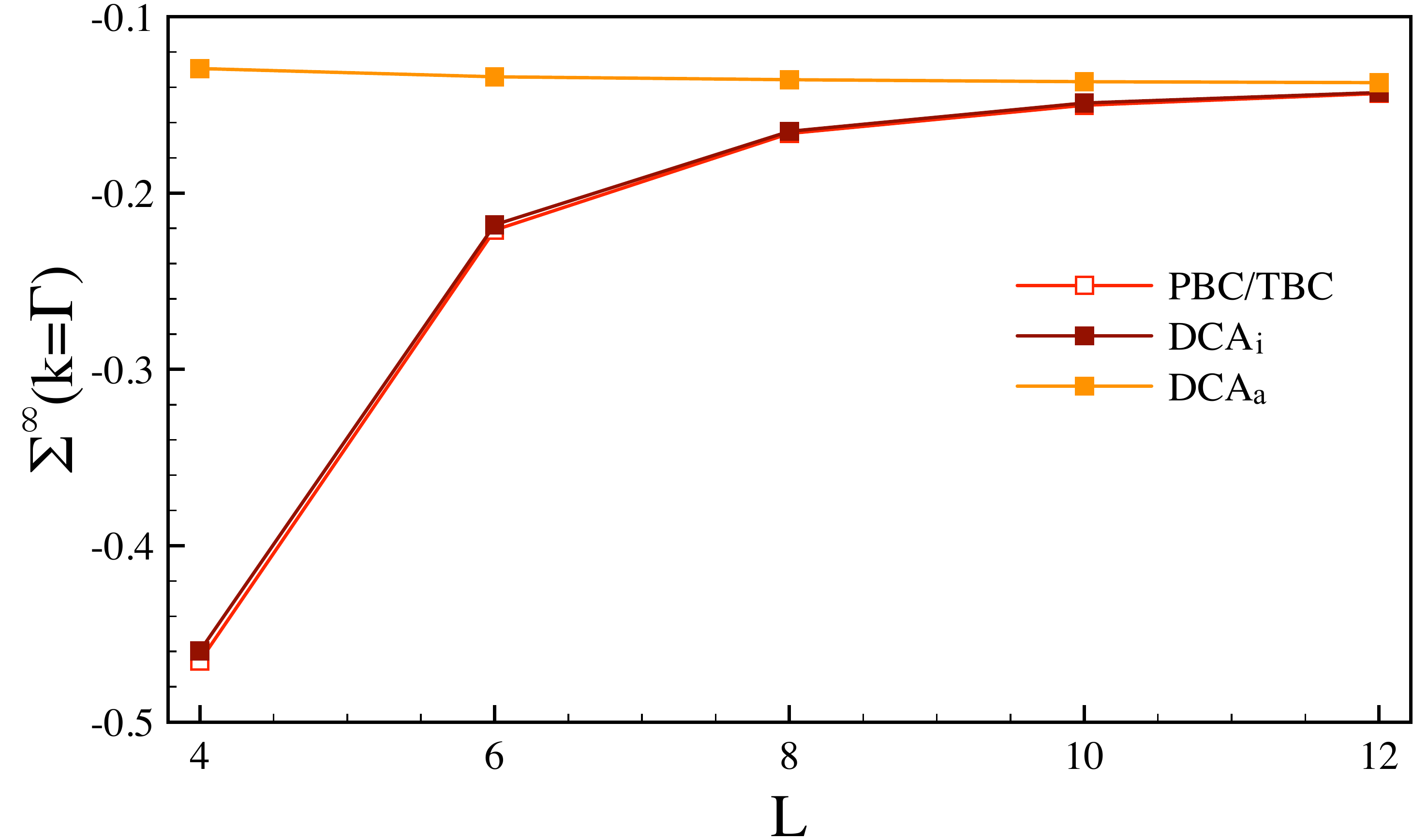}
     \caption{Static GW self-energy for the extended Hubbard model with Yukawa interaction at $U=4.0$,
         $V_0=0.2$, $\alpha=0.25$ and $\beta=6.25$ at the $\Gamma$ point as a function of system size.
         Red open squares: Clusters with periodic boundary conditions.
         Red filled squares: DCA with original interactions.
         Orange filled squares: DCA with coarse-grained interactions.
     }
     \label{fig:YU_SIGMA_Inf_U4}
 \end{figure}

Fig.~\ref{fig:YU_SIGMA_U4} shows the frequency dependence of the local Matsubara self-energy.
At the lowest
 Matsubara frequency, finite size effects persist in PBC (top left panel), while DCA with the original interactions (top right panel) achieves fast system-size convergence.
TABC results (bottom left panel) similarly show fast convergence at  low frequencies but deviations are evident at higher frequencies.
These high Matsubara frequency deviations are also evident in the DCA results with original interactions. 
 The high-frequency behavior of the Yukawa Hubbard model with longer range interactions therefore differs substantially from the extended Hubbard model with short-range interactions.
 
 In the bottom right panel of Fig.~\ref{fig:YU_SIGMA_U4}, we show DCA simulations with coarse-grained interactions
 (see Eq.~\ref{eqn:DCA_V}). We average the interaction on 27 uniformly spaced twist angles $\zeta$ for each $\tK$-patch (we cross-checked with $4\times4\times4$ twist angles at $U=4$ that $3\times3\times3$ were sufficient).
Interaction coarse-graining almost completely eliminates the high-frequency finite size errors.

 A high-frequency expansion of the
self-energy~\cite{Nolting1972,Knecht2006,Rusakov2014} shows that this high-frequency behavior is directly related to the strength of an effective
 interaction \cite{Rusakov2014}. The slow convergence at high frequency may therefore be viewed as a consequence of the approximation to the interaction.
To estimate the  finite-size convergence of these interaction finite size effects, we examine the first moment of the self-energy high frequency
 expansion~\cite{Nolting1972,Knecht2006,Rusakov2014},
 \begin{align}
     \Sigma^{(1)}_{ii} = \sum_{kl} U_{ik} U_{ki} (\langle n_k n_l\rangle - \langle n_k\rangle\langle n_l\rangle),
 \end{align}
in Table~\ref{tab:Yu_effectiveU_U4}, as a function of system-size.
Without interaction correction, PBC, TABC and DCA converge slowly with system-size.
Coarse-graining (or twist-averaging) the interaction in the DCA coarse-graining procedure significantly improves this behavior.

 \begin{table}[htb]
        \begin{tabular}{p{0.1\columnwidth}|p{0.15\columnwidth}p{0.15\columnwidth}p{0.15\columnwidth}p{0.15\columnwidth}}
         L$_c$ & \multicolumn{4}{c}{$\Sigma_1$} \\
         \hline
         & $PBC$ & $TABC$  & $DCA_1$ & $DCA_2$ \\
         \hline
         4  & 10.55 & 10.18 & 11.71 &  7.78 \\
         6  &  8.37 &  8.37 &  8.59 &  7.70 \\
         8  &  7.91 &  7.91 &  8.00 &  7.68 \\
         10 &  7.78 &    &  7.80 &  7.67
     \end{tabular}
     \caption{First high-frequency moment of the GW self-energy for the Yukawa-Hubbard model at
         $U=4$, $V_0=0.2$, $\alpha=0.25$, and $\beta=6.25$
         for various system sizes. }
     \label{tab:Yu_effectiveU_U4}
 \end{table}

 The static part of the self-energy also directly depends on interactions in the system.  Fig.~\ref{fig:YU_SIGMA_Inf_U4} shows
 convergence of the static part of GW self-energy at the $\Gamma$ point. There is a fast convergence of the
 DCA with coarse-grained interactions.

Table~\ref{tab:Yu_energy_U4} shows the convergence of the correlation
 energy for all four methods. 
For system sizes where interaction finite-size effects are not important, DCA with invariant lattice interactions
 achieves faster convergence than PBC. However, on small systems, {\it i.e.} in the presence of the strong interaction finite-size effects,
 coarse-graining is essential.

The twist-averaged correlation energy converges very fast (we also show data for $7\times7\times7$ twists). This can be understood from the fact that the
 correlation energy is dominated by the low-frequency behavior of the self-energy (high-frequency terms are suppressed by the multiplication with the Green's function, which scales $\propto 1/\omega_n$). This low-energy self-energy is  captured well by the twisted simulation (Fig.~\ref{fig:YU_SIGMA_U4}).

As the system-size is
 increased, truncation errors disappear due to the
 exponential decay of the Yukawa interaction.

 \begin{table}[htb]
        \begin{tabular}{p{0.1\columnwidth}|p{0.15\columnwidth}p{0.15\columnwidth}p{0.15\columnwidth}p{0.15\columnwidth}p{0.15\columnwidth}}
         L$_c$ & $\Delta_{\text{PBC}}$ & $\Delta_{\text{TABC}_5}$ & $\Delta_{\text{TABC}_7}$  & $\Delta_{\text{DCA}_i}$ & $\Delta_{\text{DCA}_a}$ \\
         \hline
         4  & 0.134  & 0.0167 & 0.0015 & 0.134  & 0.0112   \\
         6  & 0.0026 & 0.0021 & 0.0035 & 0.0194 & 0.0028 \\
         8  & 0.0082 & 0.0019 & 0.00012 & 0.0064 & 0.0019 \\
         10 & 0.0062 &        &         & 0.0015 & 0.000004
     \end{tabular}
     \caption{Relative correlation energy error for the Yukawa-Hubbard model at $U=4$, $V_0=0.2$, $\alpha=0.25$, and $\beta=6.25$
         for systems of size 4-10.}
     \label{tab:Yu_energy_U4}
 \end{table}

\section{Discussion and Conclusions}\label{sec:conclusions}
How should one approach the thermodynamic limit with a sequence of finite-size systems? This paper discussed two established techniques which were
developed in different communities: the technique of twist-averaging boundary conditions, and the
dynamical cluster approximation variant of cluster DMFT.

We find that TABC and DCA can be put on a similar theoretical footing, since both attempt to average over areas of the Brillouin zone that would not be considered in a naive Brillouin zone sampling.
However, the precise way of performing the average is different, as is the numerical effort. Twist-averaging is particularly efficient in Monte Carlo simulations, where the average can be taken at no additional cost during the Monte Carlo sampling. DCA is particularly efficient in semi-analytic self-consistent methods such as GW, since the DCA self-consistency can be converged during the diagrammatic self-consistency at no additional cost.

We showed that TABC provides excellent energy estimates for interactions that are contained within the finite size system. We  also showed that averaging the interactions, as is done in certain flavors of DCA, is essential if their range extends beyond the system size. Both techniques achieve far faster system size convergence than standard PBC.

Cluster DMFT results are often interpreted in terms of the historic connection to the infinite coordination number limit taken in single-site DMFT. Given the parallels between cluster DMFT and periodic  lattice calculations, we find that it is more convenient to discuss large-cluster DCA results in terms of finite-size-corrected lattice calculations, rather than in terms of `mean field' methods. This point of view  has also been taken in numerous previous works, including the large-cluster studies of two-dimensional superconductivity~\cite{Maier2005b,Gull2013} and three-dimensional Hubbard model physics~\cite{Kent2005,Gull2011a,Fuchs2011c}, and is the main reason for the success of the method in explaining the physics of the pseudogap in the two-dimensional Hubbard model~\cite{Macridin06,Werner2009,Gull2009,Gull2010}, for which recent twist-averaged calculations on small clusters~\cite{Huang2020} show remarkably similar results.

Our results suggest that averaging longer-range interactions is a promising route for controlling finite size effects, and that this technique should be attempted for long range interactions, such as the Coulomb interactions occurring in electronic structure theory.

 \acknowledgments{
     This work is supported by Simons Foundation via the Simons Collaboration on the Many Electron Problem.
     H.~T. has been supported by NSF DMR-1944974 grant.
     This research used resources of the National Energy Research Scientific Computing Center, a DOE Office of Science
     User Facility supported by the Office of Science of the U.S. Department of Energy under Contract No. DE-AC02-05CH11231
     using NERSC award BES-ERCAP0020359.
 }

 \bibliographystyle{apsrev4-2}
 \bibliography{main}

\end{document}